\documentclass[longauth]{aa} 

\usepackage{graphicx}
\usepackage{txfonts}
\usepackage{natbib}
\bibpunct{(}{)}{;}{a}{}{,} 

\usepackage{xspace}

\begin{document}

\newcommand {\x} {XMM-Newton\xspace}
\newcommand {\h} {H.E.S.S.\xspace}
\newcommand {\rchisq} {$\chi_{\nu} ^{2}$}
\newcommand {\chisq} {$\chi^{2}$}
\newcommand {\g} {$\gamma$\xspace}
\newcommand {\src} {HESS J1731$-$347\xspace}

\newcommand {\srca} {HESS J1731$-$347\xspace}
\newcommand {\srcb} {HESS J1729$-$345\xspace}

\newcommand {\radio} {G353.6-0.7\xspace}
\newcommand{\rxj} {RX~J1713.7$-$3946\xspace}
\newcommand{\vela} {RX~J0852.0$-$4622\xspace}

\newcommand {\ergs}[1]{$\times10^{#1}$ ergs cm$^{-2}$ s$^{-1}$}
\newcommand {\e}[1]{$\;\times10^{#1}$}
\newcommand {\nufnu}{$\nu F_{\nu}$ }
\newcommand {\chandra}{{\it Chandra} }
\newcommand {\nw}{nW m$^{-2}$sr$^{-1}$}
\newcommand {\micron}{$\mu$m }
\newcommand {\cms}{cm$^{-2}$s$^{-1}$ }

\newcommand {\chisphere} {28.12/6\xspace}
\newcommand {\chishell} {2.90/5\xspace}
\newcommand {\dist} {3.2\xspace}

\newcommand {\slopea} { $2.32 \pm 0.06_{\rm stat} $\xspace}
\newcommand {\norma} {$4.67 \pm 0.19_{\rm stat} $\xspace}
\newcommand {\fluxa} {$6.91 \pm 0.75_{\rm stat} $\xspace}

\newcommand {\slopeb} { $ 2.24 \pm 0.15_{\rm stat} $\xspace}
\newcommand {\normb} {$0.44 \pm 0.07_{\rm stat} $\xspace}
\newcommand {\fluxb} {$0.88 \pm 0.29_{\rm stat} $\xspace}

\newcommand {\slopesub} { $ 2.34 \pm 0.09_{\rm stat} $\xspace}
\newcommand {\normsub} {$1.41 \pm 0.11_{\rm stat} $\xspace}
\newcommand {\fluxsub} {$2.02 \pm 0.36_{\rm stat} $\xspace}

\title{A new SNR with TeV shell-type morphology: HESS J1731-347}

\author{HESS Collaboration
\and A.~Abramowski \inst{1}
\and F.~Acero \inst{2}
\and F.~Aharonian \inst{3,4,5}
\and A.G.~Akhperjanian \inst{6,5}
\and G.~Anton \inst{7}
\and A.~Balzer \inst{7}
\and A.~Barnacka \inst{8,9}
\and U.~Barres~de~Almeida \inst{10}\thanks{supported by CAPES Foundation, Ministry of Education of Brazil}
\and Y.~Becherini \inst{11,12}
\and J.~Becker \inst{13}
% ...
}

\institute{
Universit\"at Hamburg, Institut f\"ur Experimentalphysik, Luruper Chaussee 149, D 22761 Hamburg, Germany \and
Laboratoire Univers et Particules de Montpellier, Universit\'e Montpellier 2, CNRS/IN2P3,  CC 72, Place Eug\`ene Bataillon, F-34095 Montpellier Cedex 5, France \and
Max-Planck-Institut f\"ur Kernphysik, P.O. Box 103980, D 69029 Heidelberg, Germany \and
Dublin Institute for Advanced Studies, 31 Fitzwilliam Place, Dublin 2, Ireland \and
National Academy of Sciences of the Republic of Armenia, Yerevan  \and
Yerevan Physics Institute, 2 Alikhanian Brothers St., 375036 Yerevan, Armenia \and
Universit\"at Erlangen-N\"urnberg, Physikalisches Institut, Erwin-Rommel-Str. 1, D 91058 Erlangen, Germany \and
Nicolaus Copernicus Astronomical Center, ul. Bartycka 18, 00-716 Warsaw, Poland \and
CEA Saclay, DSM/IRFU, F-91191 Gif-Sur-Yvette Cedex, France \and
University of Durham, Department of Physics, South Road, Durham DH1 3LE, U.K. \and
Astroparticule et Cosmologie (APC), CNRS, Universit\'{e} Paris 7 Denis Diderot, 10, rue Alice Domon et L\'{e}onie Duquet, F-75205 Paris Cedex 13, France \thanks{(UMR 7164: CNRS, Universit\'e Paris VII, CEA, Observatoire de Paris)} \and
Laboratoire Leprince-Ringuet, Ecole Polytechnique, CNRS/IN2P3, F-91128 Palaiseau, France \and
Institut f\"ur Theoretische Physik, Lehrstuhl IV: Weltraum und Astrophysik, Ruhr-Universit\"at Bochum, D 44780 Bochum, Germany
% ...
}

%
\offprints{facero@in2p3.fr}
   \date{Received 2010 December 31; accepted 2011 May 6}
%
%
  \abstract
   {}
   {The recent discovery of  the radio shell-type supernova remnant (SNR), \object{G353.6-0.7}, in spatial coincidence with the unidentified TeV source \src has motivated further observations
   of the source with the High Energy Stereoscopic System (\h) Cherenkov telescope array to test a possible association of the \g-ray emission with the SNR.}
   {With a total of 59 hours of observation, representing about four times the initial exposure available in the discovery paper of \object{HESS~J1731$-$347}, the \g-ray morphology 
   is investigated and compared with the radio morphology. An estimate of the distance is derived by comparing the interstellar absorption derived from 
    X-rays and the one obtained from $^{12}$CO and HI observations. } 
   {The deeper \g-ray observation of the source has revealed 
   a large shell-type structure  with similar position and extension (r$\sim$0.25$^{\circ}$) as the radio SNR, thus confirming their association.
   By accounting for the \h angular resolution and projection effects within a simple shell model, the radial profile is compatible with a 
   thin, spatially unresolved, rim. 
   Together with \rxj, \vela and SN 1006, \src is now the fourth SNR with a significant shell morphology at TeV energies. 
    The derived lower limit on the distance of the SNR of \dist kpc is used together with radio and X-ray data 
    to discuss the possible origin of the \g-ray emission, either via inverse Compton scattering of electrons or the decay of neutral pions 
    resulting from proton-proton interaction.   }
   {}

   \keywords{  Astroparticle physics - Gamma-rays : observations -  ISM: supernova remnants -     SNR : individual : \srca        }

\authorrunning{The HESS Collaboration}
\titlerunning{HESS~J1731$-$347 a new TeV shell-type SNR}

   \maketitle

\section{Introduction}

In the survey of the Galactic plane carried out by the \h experiment, many sources emitting at TeV energies remain unidentified to date  \citep[e.g.][]{ah08}.
Most of the sources are extended beyond the point spread function (PSF) of the  \h experiment ($\sim 0.06^{\circ}$ for the analysis presented in this paper). The largest number of conclusive 
identifications so far can be attributed to pulsar wind nebulae (PWNe) as presented in e.g. \citet{gallant08}. 
Recently, a new radio SNR, catalogued as \radio, was discovered by \citet{tl08} to be in spatial coincidence with \src,  one of the
 unidentified sources presented in \citet{ah08}. The diameter of the radio shell is nearly 0.5$\degr$ 
which allows, given the brightness of the source and the \h angular resolution of $\sim$0.06$\degr$, for a morphological comparison of the \g-ray source with the
 shell observed in radio. 
Moreover, at least up to the current date, no radio pulsar or X-ray PWN candidate was found that might alternatively explain the TeV emission. 
This situation should be compared to other \g-ray sources like HESS\,J1813$-$178 or HESS\,J1640$-$465 \citep{aharonian06}, which are also in spatial coincidence with radio SNR shells. However, for these latter sources a morphological identification with the radio shells is not possible, 
and the emission can plausibly originate from a PWN seen in X-rays as discussed in
\citet{funk07b} for HESS\,J1640$-$465 and in \citet{funk07,gotthelf09} for   HESS\,J1813$-$178.

Observations  of the north-eastern part of  \src with the X-ray satellites \x, Chandra, and Suzaku have confirmed an X-ray counterpart
 found in archival ROSAT data \citep[presented in ][]{ah08,tl08}.
An X-ray shell partly matching
the radio morphology was found and the spectral analysis has revealed that the X-ray emission is of 
synchrotron origin,  indicating  that the shock
wave of the SNR has accelerated electrons up to   TeV  energies \citep{ap09,tl10}.
A compact (unresolved) X-ray source XMMU J173203.3$-$344518  \citep{halpern10} was observed towards the geometrical center of the remnant
and has spectral properties reminiscent of central compact objects (CCOs) found in several other supernova shells  \citep[e.g.][]{ps04}.
A search for pulsations using the EPIC PN cameras onboard \x shows only marginal evidence of a 1\,s period \citep{halpern10}.

Given the recent discovery of \radio, little is known about its age and distance. \cite{tl08} suggested a distance of 3.2 $\pm$ 0.8 kpc  
assuming that the SNR is at the same distance as the H{\sc ii} region G353.42-0.37.

Additional \h observations, carried out since the discovery paper of \src \citep{ah08}, allow  to investigate the compatibility of the
TeV source with the radio shell SNR \radio.  The observations and the data analysis are described
in Sect. \ref{sect:data} and  the morphological and spectral results in Sect. \ref{sect:results}. 
The multi-wavelength counterparts of \src are described in Sect.  \ref{sect:multi} and a general discussion is presented in Sect. \ref{sect:discussion}.

\section{\h observations and analysis methods }
\label{sect:data}

\h  is an array of four identical imaging atmospheric Cherenkov telescopes (IACTs)
located in the Khomas Highland of Namibia 1800 m above sea level   \citep{bernlohr03}.
The survey of the Galactic plane by the \h collaboration has led to the discovery of 
the \g-ray source HESS J1731-347, presented as an unidentified extended source in \citet{ah08}.
In this first data set,  14 hours of observation time were available. 
Additional dedicated observations were carried out in July 2007 and 
in July and August 2009 with zenith angles ranging from 9$\degr$ to 42$\degr$, the mean angle being 16.5$\degr$. 
The total \h observation time for this target is 59 hours after data quality cuts.

The  data set
was analyzed using the \textit{Model analysis} \citep{dr09} which exploits the
full pixel information by comparing the recorded shower images with a pre-calculated shower model
using log-likelihood minimization. In comparison with conventional analysis techniques, no cleaning  or parametrization of the image shape is required and the full camera information is used.
This method leads to a more accurate reconstruction and better background suppression
than more conventional techniques and thus to an improved sensitivity.

Spectral and spatial analyses were carried out using a minimum image intensity of 60 photoelectrons (p.e.) resulting in an energy threshold of 240 GeV and an angular resolution of  0.06$^{\circ}$ (68\% containment radius).
 All results presented were cross-checked with a multivariate analysis \citep{ohm09} using an independent calibration and gamma/hadron separation,
  which yielded consistent results. Unless otherwise quoted, the error bars in the following section are given at 1$\sigma$. \\
 
\section{TeV \g-rays analysis results}
\label{sect:results}

\begin{figure}
   \centering

   \includegraphics[bb= 89 200 465 565, clip,width=7.7cm]{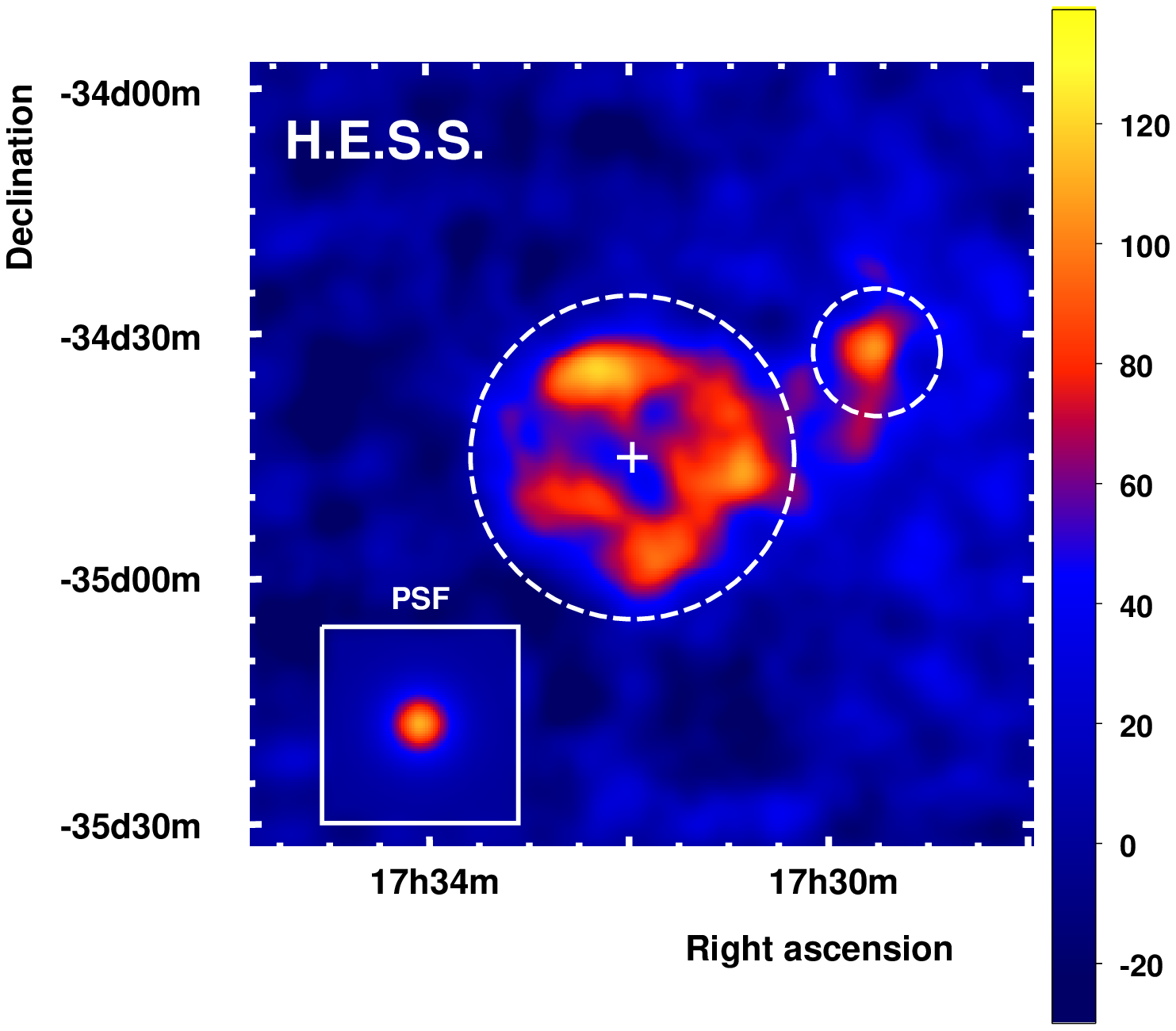}   
   \includegraphics[bb= 465 114 516 581, clip,width=0.812cm]{16425f1.ps}
   
      \vspace{-0.6cm}

   \caption{TeV \g-ray excess map ($1.5\degr \times1.5\degr$) of the \src region smoothed with a Gaussian width $\sigma$=0.04$^{\circ}$.
    The average \h PSF for the dataset is shown in the inset. The regions used for the spectral analysis of \srca and \srcb are respectively represented by the large and small dashed circles.
    The position of the central compact object detected in X-rays is shown with a white cross. The linear scale is in units of excess counts per smoothing Gaussian width. The transition between blue and red in the color scale is at the level of 4$\sigma$.}
   \label{fig:excess}
\end{figure}

The \h  excess map of the region of \src is shown in Fig.~\ref{fig:excess} smoothed with a Gaussian of $\sigma$=0.04$^{\circ}$. 
For the background estimation in the image and in the morphology studies, the \textit{ring background} method presented in \citet{berge07} was used.
Because of the larger data set and the more sensitive reconstruction technique, 
the presented image is much more detailed than the one shown in the
discovery paper \citep{ah08}. 
This reveals a complex region composed of a large and bright structure (\srca), detected
at 22$\sigma$, with a suggestive shell-like morphology.

A smaller and fainter structure named \object{\srcb} (detected at 8$\sigma$) is also observed, the 
properties of which are presented separately in Sect. \ref{sect:srcb}.

\subsection{TeV energy morphology}

To further test    the hypothesis of a shell morphology for \srca and its association with the radio SNR, radial and azimuthal profiles in radio and \g-rays were extracted centered on the
position of the CCO ($\alpha_{\mathrm{J2000}}=$17h32m03s,  $\delta_{\mathrm{J2000}}=-34\degr45'18''$), also coincident
with the geometrical center of the radio SNR. The profiles were derived from the uncorrelated \g-ray excess map
and corrected for the field of view (FoV) acceptance.
  
For the  \g-ray radial profile, the position angles\footnote{Position angle 0$\degr$ corresponds to North and 90$\degr$ to East.} from 270$\degr$ to 310$\degr$ were excluded, to avoid contamination from \srcb, and the resulting radial profile was compared with a sphere and a shell model. 
The first model is a uniformly emitting sphere of adjustable radius, projected on the sky and then folded with  the PSF derived
for this analysis ($r_{\rm 68\%}=0.06\degr$).
The shell model consists of a uniformly emitting shell of variable outer radius and thickness (defined as $r_{\rm outer} - r_{\rm inner}$) projected on the sky and then folded with the same PSF. 

In the morphological test, the best fit statistically favors the shell model and the sphere model  is 
ruled out at 3.9 $\sigma$ ($\chi^{2}$/dof = \chishell  and \chisphere   
for the shell and sphere models respectively).  
In the case of a shell model, the best fit  radius is $0.27^{\circ}  \pm \,0.02^{\circ}$ and the emission is compatible with  a thin, spatially
unresolved, shell with an upper limit thickness of 0.12$^{\circ}$ (90\% confidence level). 
 
\begin{figure}
   \centering
      \includegraphics[width=\columnwidth]{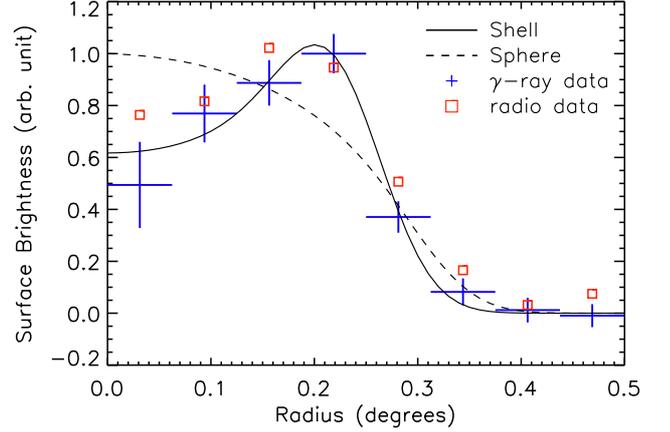}
            \vspace{-0.6cm}
   \caption{The \g-ray excess and radio radial profiles are shown with  green crosses and red squares respectively. 
The best fits  to the \g-ray data of a sphere and a shell model are overlaid. Both radial profiles 
are centered on the compact central object ($\alpha_{\mathrm{J2000}}=$17h32m03s,  $\delta_{\mathrm{J2000}}=-34\degr45'18''$). }
   \label{fig:radprof}
\end{figure}

To compare the TeV morphology with the shell seen in radio, 
 the radio continuum map from the ATCA southern Galactic plane survey (SGPS)  \citep{hg06} was 
smoothed to match the \h spatial resolution and a radial profile was extracted (excluding point sources).
The radio profile was then scaled  by a normalization factor
calculated as the ratio of the total number of excess  $\gamma$-rays over the total radio flux  on the whole remnant. 
The resulting profiles, presented in Fig.~\ref{fig:radprof}, show an extended emission in 
\g-rays similar to that seen in radio.

\begin{figure}
   \centering
   \includegraphics[width=\columnwidth]{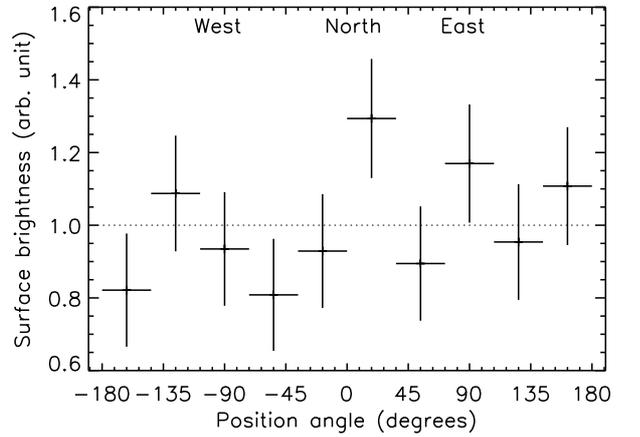}
      \vspace{-0.6cm}
   \caption{Normalized azimuthal \g-ray excess profile restricted to radius r $\leq0.3\degr$ and using the same center as in Fig.~\ref{fig:radprof}. The brightness distribution is compatible with a flat profile.}
   \label{fig:azprof}
\end{figure}

In contrast with \rxj which is brighter in the North-West and SN 1006 that exhibits a bipolar morphology, 
the  azimuthal profile of \srca (see Fig.~\ref{fig:azprof}) integrated   for $r \leqslant 0.3 \degr$   shows no significant deviation from a flat profile ($\chi^{2}$/dof = 8.8 / 9).

\subsection{Spectral results}

\begin{figure}
   \centering
   \includegraphics[width=\columnwidth]{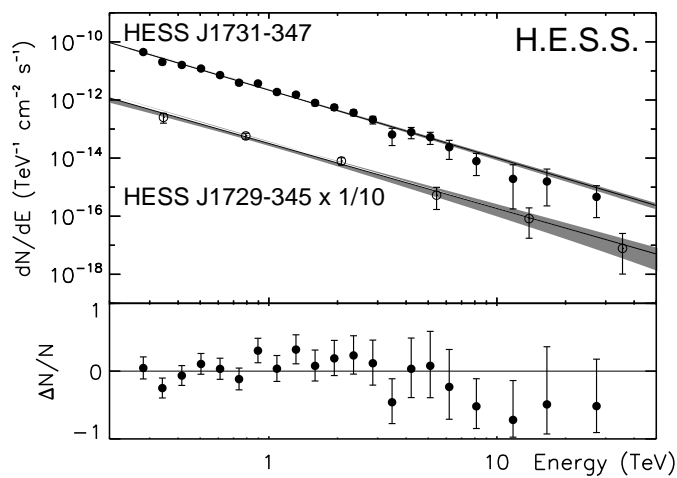}
   \caption{Differential energy spectra of \srca (filled circles) and \srcb (open circles). The normalization for the second source has been divided by 
   10 for graphical purposes.  Events were binned to reach a significance of at least 2$\sigma$. 
   The best fit power-law models along with the residuals for \srca are also shown.
   The grey bands correspond to the range of the power-law fit, taking into account statistical errors.}
   \label{fig:spec}
\end{figure}

The energy spectrum of the SNR was obtained by means of a forward-folding maximum likelihood fit \citep{piron01}
from a circular region of 0.3$\degr$ centered 
on the CCO, illustrated by the large dashed circle ($r=0.3^{\circ}$) in Fig.~\ref{fig:excess}, chosen to fully enclose
the emission of the remnant. 
The background is estimated using the \textit{multiple reflected-regions}
technique where background events are selected from regions of the
same size and shape as the source region and at equal angular distance from the
observation position \citep{berge07}. 
The resulting spectrum, shown in Fig.~\ref{fig:spec},   is well described by  a power-law model (equivalent $\chi^{2}$/dof = 27.7 / 35) defined as
${\rm d}N/{\rm d}E = N_{\rm 0} (E/E_{\rm 0})^{-\Gamma} $ where $E_{\rm 0}$ is the decorrelation energy (energy at which the correlation between the slope and the normalization vanishes).
 The best fit parameters, listed in Table \ref{tab:param}, 
result in an integrated 1-10 TeV energy flux of (\fluxa $\pm \, 1.38_{\rm syst}) \times 10^{-12}$ erg cm$^{-2}$s$^{-1}$.  The flux measured here is lower than what has been derived initially in
 \citet{ah08} : $(16.2 \pm  3.6_{\rm stat} \pm \, 3.2_{\rm syst}) \times 10^{-12}$ erg cm$^{-2}$s$^{-1}$ in the same energy band.
  However, the region of extraction in the discovery paper was much larger ($r=0.6^{\circ}$ versus $r=0.3^{\circ}$
 in this paper),  including \srcb and possibly some surrounding diffuse emission. 
 A cross-check to derive the flux from the SNR only using the same data set as used in \citet{ah08} and following the original analysis method 
 gave results consistent with the complete data set presented here thus confirming that the flux difference was mainly due to the choice of the integration region.
 A power-law model with an exponential cutoff was also tested which did not improve the quality of the fit (equivalent  $\chi^{2}$/dof = 24.0 / 34).

\begin{figure*}[!t]
   \centering
   \begin{tabular}{ccc}
\hspace{-0.6cm}
 { \includegraphics[bb= 43 137 570 650 , clip,width=6.23cm]{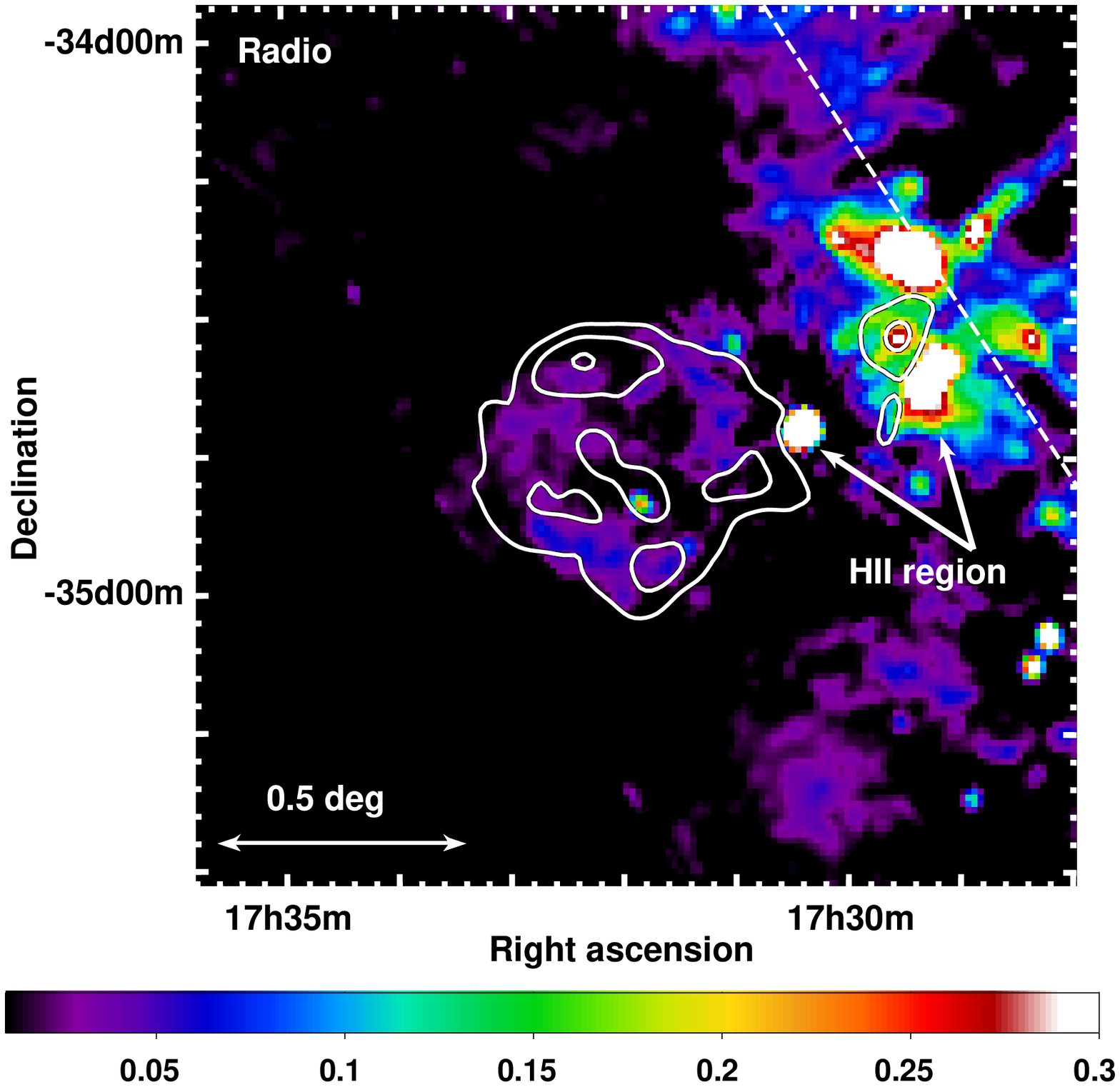}  } & \hspace{-6mm}
 { \includegraphics[bb= 40 137 565 641 , clip,width=6.20cm]{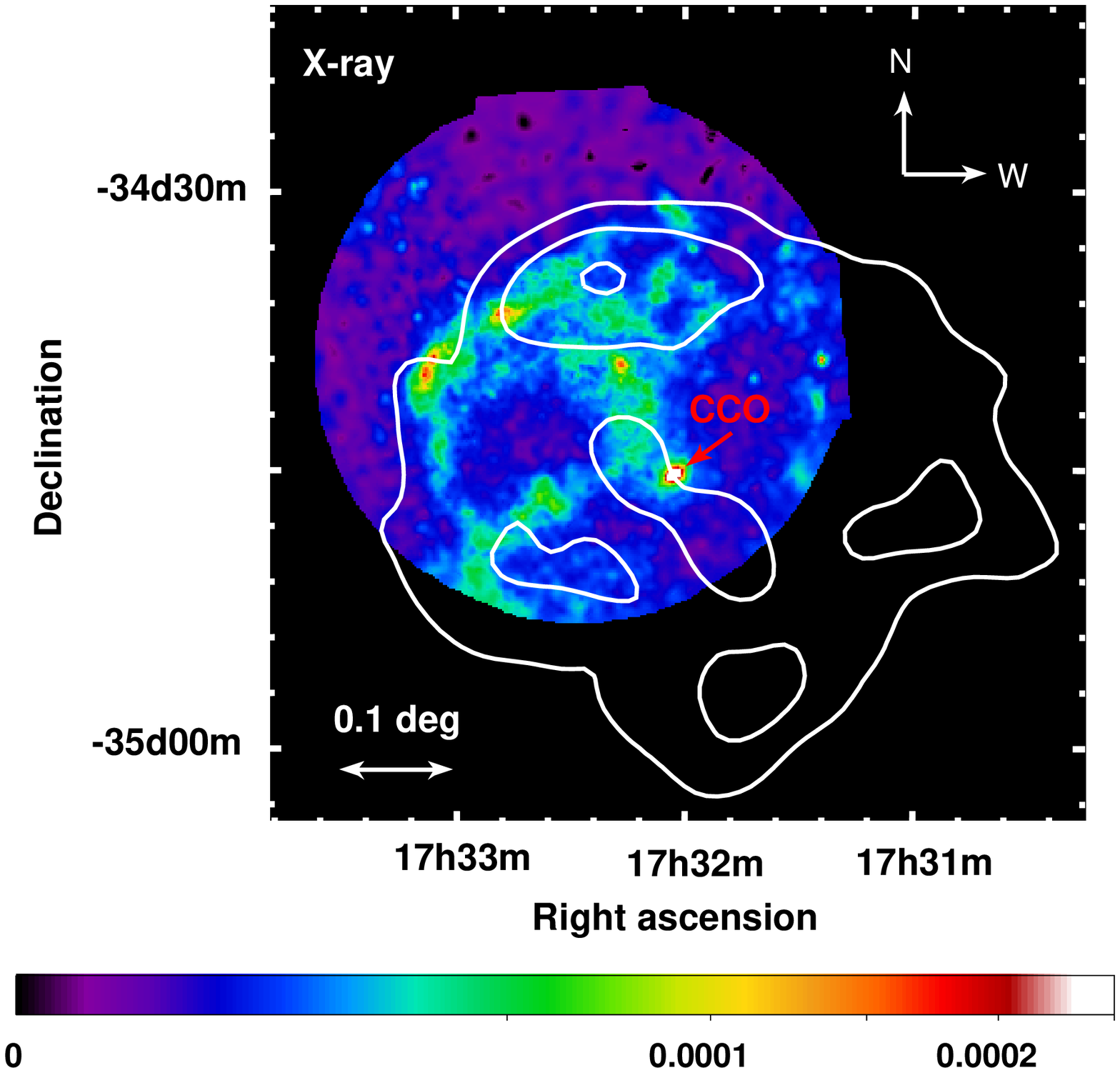} }  & \hspace{-6mm}
 { \includegraphics[bb= 43 137 570 648 , clip,width=6.23cm]{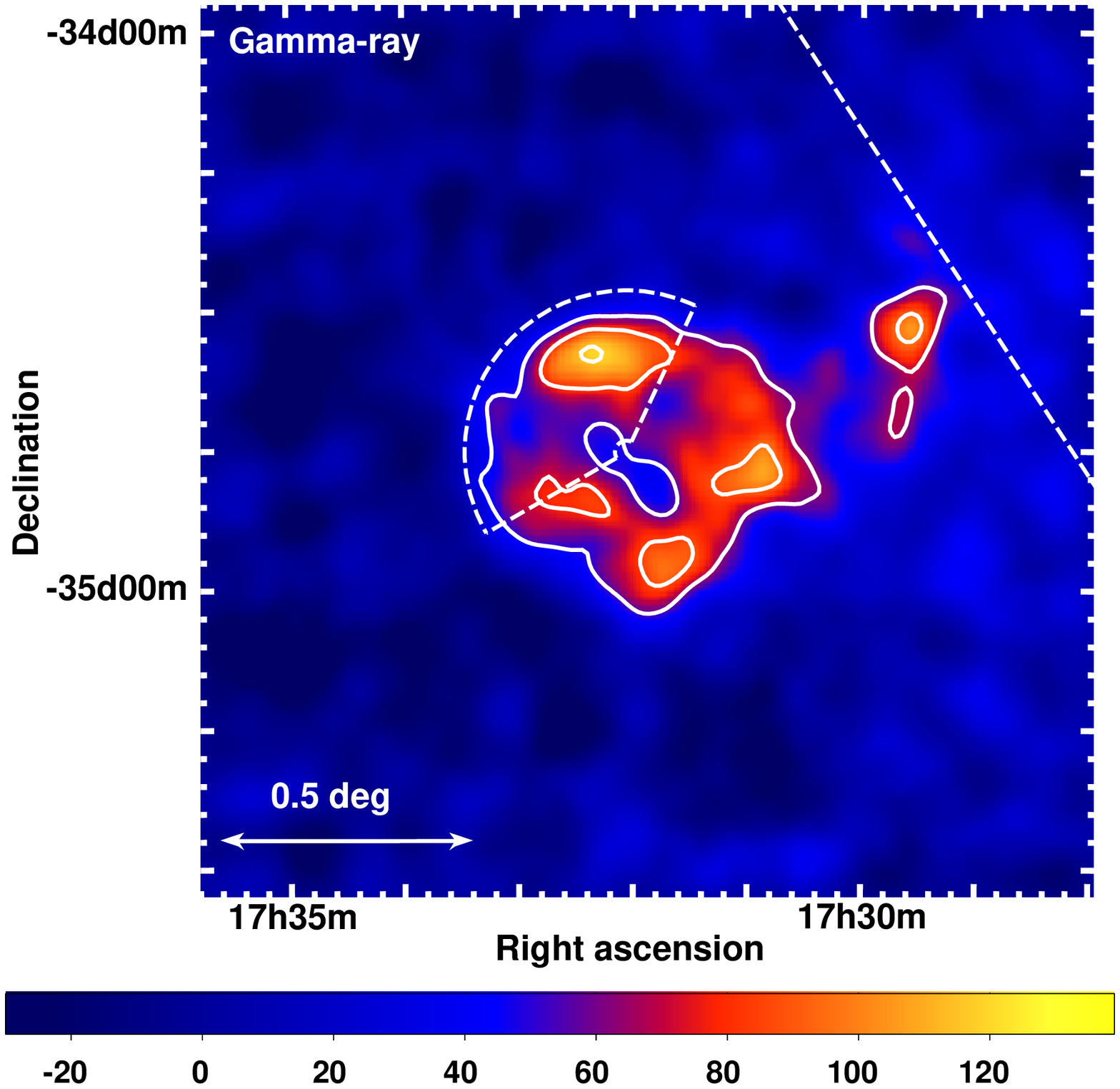} } 
 
   \end{tabular}

\vspace{-3mm}

   \caption{Multi-wavelength view of the  \src region. The radio and the \g-ray image show the same field
    of view  while the X-ray image is zoomed in order to show the details of the shell structure. The significance contours at 4, 6 and 8 $\sigma$ obtained with an
   integration radius of 0.06$\degr$ and the Galactic plane (white dashed line) are overlaid in the three panels.  
     \textit{Left} : ATCA radio map at 1.4 GHz from the south Galactic plane survey (SGPS)
     in units of Jy/beam with a beam of 100''. 
   The HII regions G353.42-0.37 (left) and G353.381-0.114 (right) are marked with arrows. \textit{Middle} :   \x observation of a sub-region of the SNR, in the 0.5-4.5 keV energy band, using MOS instruments with units in ph/cm$^{2}$/s/arcmin$^{2}$.  The position of the source XMMU J173203.3$-$344518, 
   which is likely to be the CCO of the SNR is shown by the red arrow. \textit{Right} : 
   TeV \g-ray excess map of \src  smoothed with a Gaussian with $\sigma$=0.04$^{\circ}$.
    The region used to derive the radio flux and the spectral parameters in X- and \g-rays for the SED is also shown. }

   \label{fig:4contrib}
\end{figure*}

\subsection{\srcb}
\label{sect:srcb}

A \g-ray excess of TeV emission was found  at the best fit position 
$\alpha_{\mathrm{J2000}}=$17h29m35s,  $\delta_{\mathrm{J2000}}=-34\degr32'22''$ with a statistical error of 0.035\degr and the source was therefore
labeled \srcb.

The source is extended beyond the size of the PSF  (Gaussian width $\sigma=0.12\degr \pm 0.03\degr$)
 and the region used to derive 
the spectral parameters is shown by the small dashed circle ($r=0.14\degr$) in Fig.~\ref{fig:excess}. The spectrum obtained is well modeled by a power-law model (see Fig.~\ref{fig:spec}) and the best fit parameters are listed in Table \ref{tab:param}. The integrated flux in the 1-10 TeV energy band is  (\fluxb $\pm \,   0.18_{\rm syst}) \times 10^{-12}$ erg cm$^{-2}$s$^{-1}$.

\begin{table*}
\begin{minipage}[t]{\textwidth}
\centering
\caption{Best fit spectral parameters obtained for different extraction regions in \src. The model used is a power-law of the form ${\rm d}N/{\rm d}E = N_{\rm 0} (E/E_{\rm 0})^{-\Gamma} $. The systematic errors are conservatively  estimated  to be $\pm$ 0.2 on the photon index and 20\% on the flux.   }
\label{tab:param}

\renewcommand{\footnoterule}{}  
\begin{tabular}{l c c c c  }     
\hline

Region & Photon index $\Gamma$ & Decorrelation energy $E_{\rm 0}$& Normalization $N_{\rm 0}$ & 1-10 TeV integrated flux \\
           &                          & TeV            &   $10^{-12}$ cm$^{-2}$s$^{-1}$ TeV$^{-1}$  &  $10^{-12}$ erg cm$^{-2}$s$^{-1}$ \\
\hline

   \srca & \slopea &  0.783 & \norma & \fluxa  \\
   sub-region of  \srca \footnote{A spectral analysis corresponding to the FoV of the \x  data (see Fig.~\ref{fig:4contrib}, center) has been carried out in order to build a SED.}  &   \slopesub &  0.780 &\normsub & \fluxsub  \\
      \srcb &   \slopeb & 0.861 &\normb & \fluxb  \\

\hline
\end{tabular}
\end{minipage}
\end{table*}

\section{Multi-wavelength counterparts}
\label{sect:multi}

One of the interesting characteristics  of \srca is that  non-thermal emission is clearly identified in 
radio, X-rays and at  TeV energies.
In X-rays however,  the access to the spectral properties  is limited to a subregion of the SNR as 
the coverage with the \x, Chandra and Suzaku  satellites is only partial, and the statistics in the ROSAT All Sky Survey data are too low.
In order to study the spectral energy distribution (SED) of the source, 
 the radio flux and the TeV spectral properties were extracted only from the region observed in X-rays (see region definition in Fig.~\ref{fig:4contrib}, right).  
The multi-wavelength counterparts of \srcb are discussed later in Sect. \ref{sect:robin}.

\subsection{Radio Continuum}
\label{sect:radio}

The shell  observed in radio is spatially coincident with the \g-ray shell and has a similar extent (see radial profile in Fig.~\ref{fig:radprof}).
The flux obtained (excluding point sources)  from the SGPS ATCA data in the region observed by the \x pointing is  0.8 $\pm$ 0.3 Jy at 1420 MHz.
The total radio flux for the SNR measured by  \citet{tl08} is of 2.2 $\pm$ 0.9 Jy.
The compact HII region (G353.42-0.37) located to the West of the remnant at a distance of  3.2 $\pm$ 0.8 kpc \citep{tl08} is indicated 
in Fig.~\ref{fig:4contrib} (left).

\subsection{X-rays}
\label{sect:xray}

In order to derive spectral information from the X-ray emission from the remnant, the \x pointing obtained as a follow up of the HESS source 
(ObsId: 0405680201 ; PI: G. P\"uhlhofer) was analyzed. 
To clean the proton flare contamination during the observation, a histogram of the 10-12 keV count rates of each 
camera was built. A Gaussian fit was then performed in order to remove time intervals where the count rates were beyond 3 $\sigma$ from the mean value \citep{pa02}. The remaining exposure time after flare screening is 22 ks out of the 25 ks of observation for MOS and 15 ks for PN. For the image generation, the instrumental background was derived from the compilation of blank sky observations by \citet{cr07}
 and renormalized in the 10-12 keV band over the whole  FoV. The image resulting from the combination of the
  two MOS instruments  is presented in Fig.~\ref{fig:4contrib} (middle). 
For this mosaic, the data from the PN instrument were not used because of  straylight contamination to the
North-East (photons  singly reflected by the mirrors) from a bright X-ray source located outside the FoV. 
This results in some spurious arc features near the border of the FoV in the North-East.

The X-ray emission is characterized by extended emission which is concentrated in arc-like features, similar to broken shell seen from many shell-type SNRs. 
Some of the arcs partly coincide with the radio and \g-ray shell (see Fig.~\ref{fig:4contrib}). Some of the structures could hint at an additional, smaller shell, 
but might also come from irregular SNR expansion in an inhomogeneous and/or dense medium \citep{blondin01}.  
A double-shell structure is also observed in \rxj in  X-rays \citep{lazendic04,cd04,acero09}. 

The spectral analysis of the diffuse X-ray emission was carried out using the  Extended Source Analysis Software
(ESAS\footnote{http://xmm2.esac.esa.int/external/xmm\_sw\_cal/background/
epic\_esas.shtml})
 provided in the \x Science Analysis System (SAS v9.0) to model the particle and instrumental backgrounds. 
The error bars in this section are quoted at 90\% level confidence.
For this analysis, the three instruments PN+MOS1+MOS2 were used and the regions were selected to avoid the straylight features.

\begin{figure}
   \centering
   \includegraphics[bb= 70 25 590 705,clip,width=6.4cm,angle=-90]{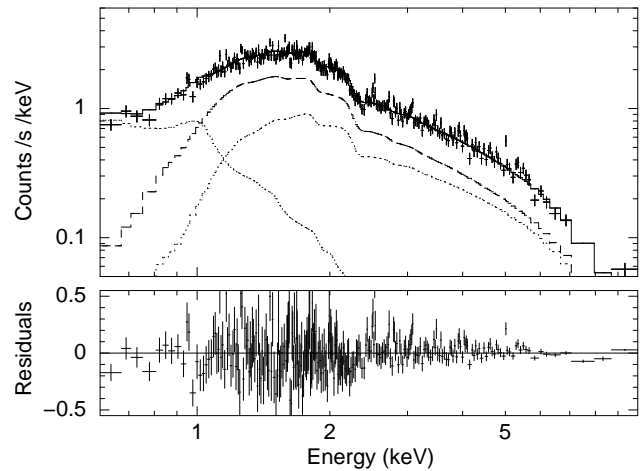}
   \vspace{-0.2cm}

   \caption{X-ray spectrum, using the PN camera, extracted from the sub-region of \srca shown in Fig.~\ref{fig:4contrib} (right). 
   The non-thermal emission from the SNR  is  described by an absorbed power-law model (dashed line). 
   The local astrophysical background, fitted to an off region outside the SNR, is modeled by two components (dotted lines).
    The low energy component is an APEC model (astrophysical plasma emission code, see http://hea-www.harvard.edu/APEC)
  representing the background from the Local Bubble and
     the high-energy component is an absorbed power-law representing the hard X-ray background  (unresolved AGNs, cataclysmic variables, etc). 
     The residuals of the total model (SNR+local astrophysical background) are shown in the lower panel and the $\chi^{2}$/ndof  is 1921/1569.}
   \label{fig:xrayspec}
\end{figure}

\begin{figure*}
   \centering
    \begin{tabular}{cc}
    
   {\includegraphics[bb=-80 28 504 519,clip,width=7.8cm]{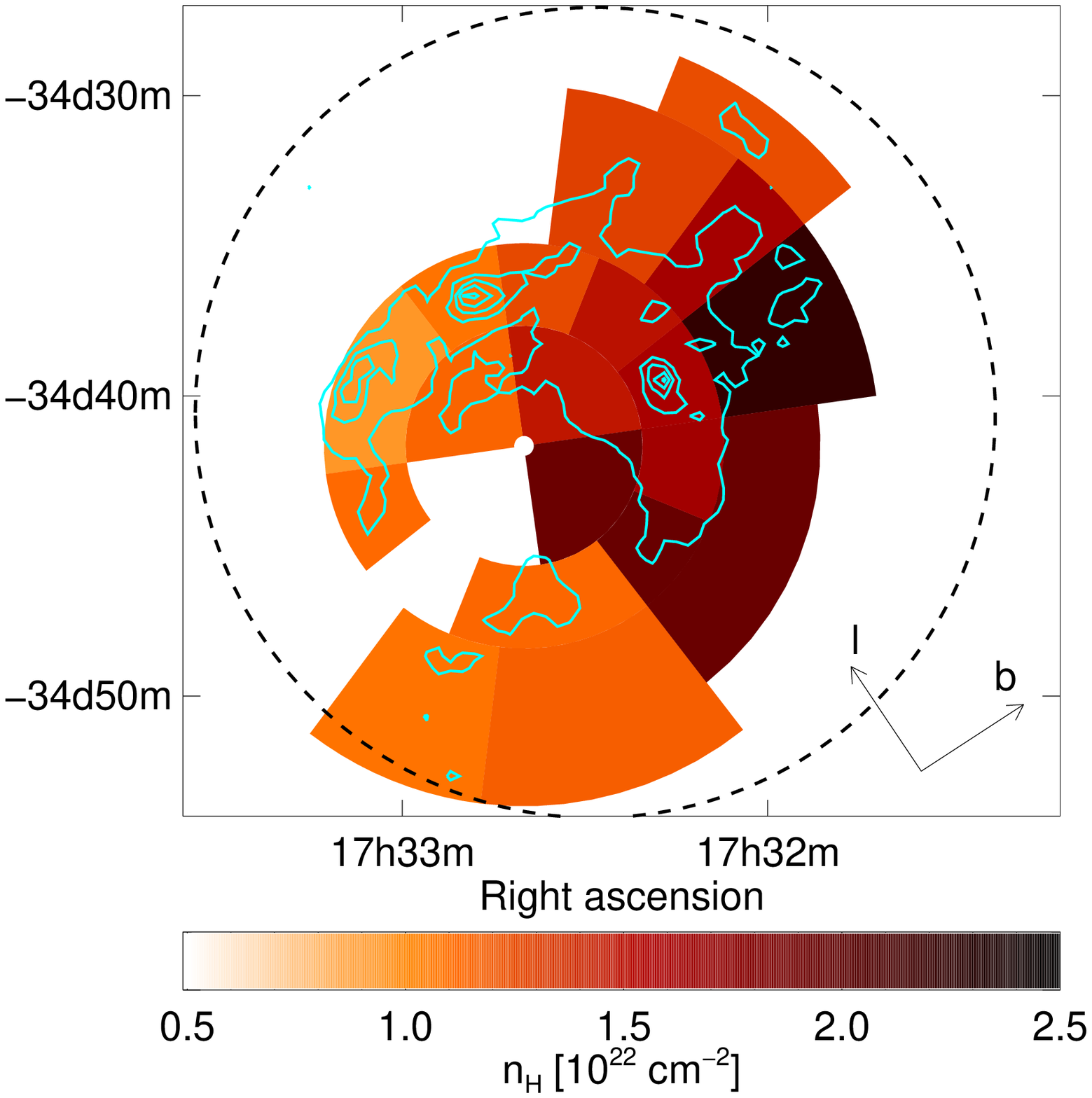} } & \hspace{1.cm}
   {\includegraphics[bb=45 156 570 620,clip,width=7.6cm]{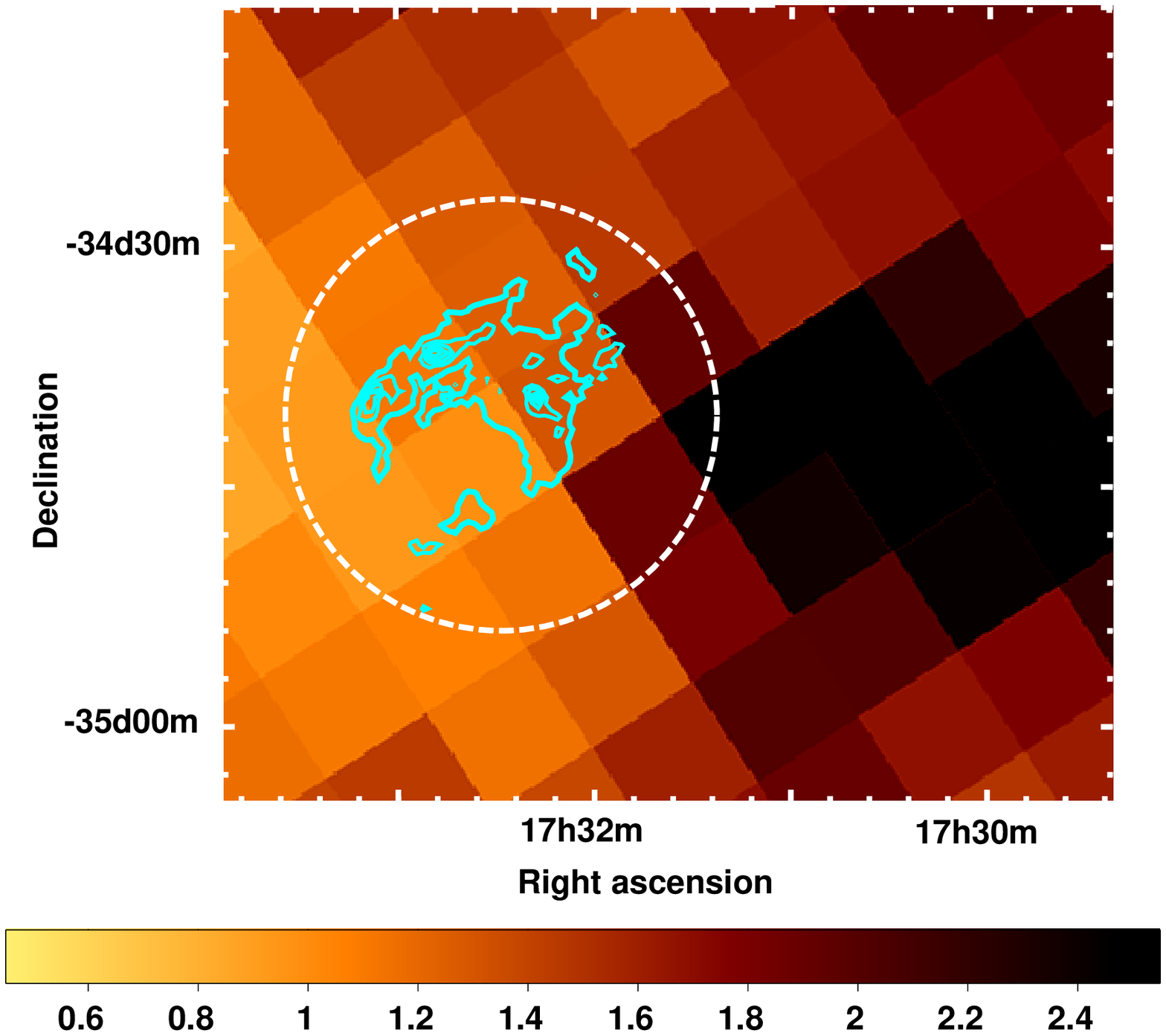} } \\   \vspace{0.5cm}
   
 \hspace{-2.1cm}   {\includegraphics[bb=-120 0 440 27,clip,width=7.5cm]{16425f9.ps}}      &   
  \hspace{-1.5cm} {\includegraphics[bb=-120 0 440 27,clip,width=7.5cm]{16425f9.ps}} 
   
      \vspace{-0.7cm}

   \end{tabular}

   \caption{ \textit{Left } : X-ray absorption map derived from a spectral fit to \x data assuming an absorbed power-law model. A significant increase of $N_{\rm H}$ towards the Galactic plane is observed. \textit{ Right } : Absorption column map derived from atomic and molecular hydrogen when integrating over radial velocities from 0 km/s to  $-$25 km/s (see Sect. \ref{sect:CO}  for more details). The Galactic plane is represented by the white dashed line. In both panels, the \x field of view is represented by a dashed circle and the X-ray contours
   obtained from Fig.~\ref{fig:4contrib} (center)  are overlaid. }
   \label{fig:xraynh}
\end{figure*}

The spectrum derived from the region covered by the FoV of \x  that is used for the SED  is shown in Fig.~\ref{fig:xrayspec}. 
The emission is well represented by an absorbed power-law model and no emission lines were found \citep[see also][]{tl10}. 
The best fit parameters obtained from  a joint fit of MOS1, MOS2 and PN spectra  are 
$N_{\rm H}=(1.08 \pm 0.02) \times 10^{22}$ cm$^{-2}$,  a spectral index $\Gamma= 2.28  \pm 0.03$ and a normalization at 1 keV  
$N_{\rm 0}= (1.37  \pm 0.05) \times 10^{-2}$ cm$^{-2}$  s$^{-1}$ keV$^{-1}$.
A search for spatial spectral variations  of the diffuse emission revealed that the power-law index is in most locations in the 
range $\Gamma = 2.1-2.5$. Under the assumption  of a pure power-law hypothesis, the  absorption column significantly increases towards the Galactic plane from 
$N_{\mathrm{H}} = (0.93  \pm  0.05) \times 10^{22}$ cm$^{-2}$ in the South-East region to 
$N_{\mathrm{H}} = (2.23   \pm 0.21) \times 10^{22}$ cm$^{-2}$  in the North-West region (see Fig.~\ref{fig:xraynh} ; left).
The errors on the absorption column shown in Fig.~\ref{fig:xraynh} (left) are ranging from 5\% to 12\%.

 The bright point source XMMU J173203.3$-$344518 lies at the geometrical center of the radio and \g-ray shell.  
Marginal evidence for a pulsation at a period of 1 s for the pulsar candidate and a faint nebula (radius of 30'') 
whose spectral properties are compatible with a dust-scattered halo have been reported by \citet{halpern10}. 
As no optical or 
IR counterpart of the point source have been  detected and as the X-ray spectrum is well described by a blackbody emission model 
with $kT\sim0.5$ keV,  the object is
a good candidate to be  the CCO of the SNR \citep{ap09,halpern10,tl10}.

\subsection{$^{12}$CO (J=1-0) and HI}
\label{sect:CO}

\begin{figure}
   \centering
   \includegraphics[bb=70 360 585 1040,clip,width=\columnwidth]{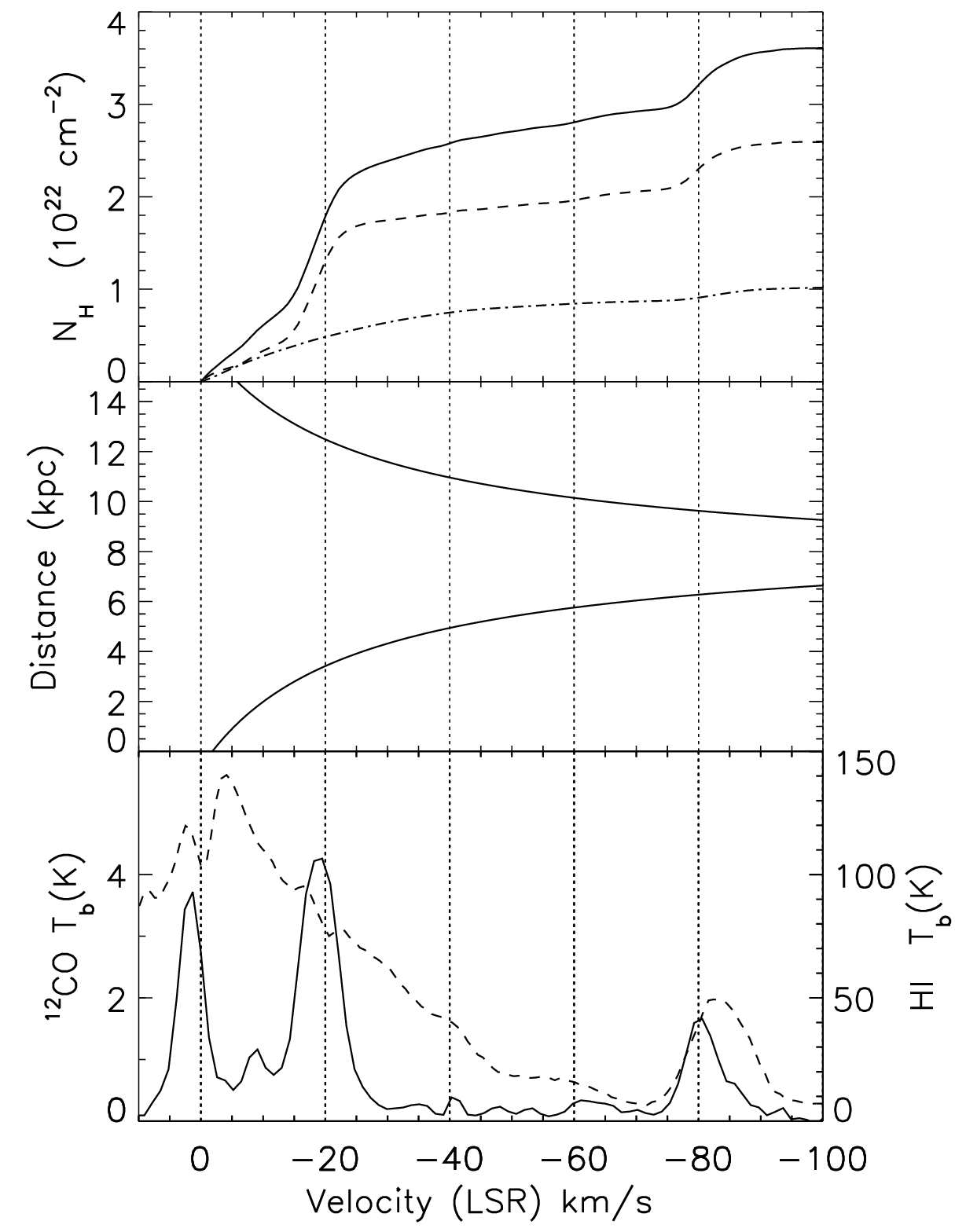}
       \vspace{-0.5cm}
     \caption{ 
      \textit{Top} : Cumulative absorbing column density (solid line) as a function of radial velocity at the position of highest X-ray absorption (see Sect. \ref{sect:CO}).
       The relative contributions from the atomic and molecular hydrogen are represented by the dashed and dash-dotted lines respectively.
   \textit{Middle} : Rotation curve towards the same direction as derived from the model of Galactic rotation of  \citet{hou09}.   
    \textit{Bottom} :  $^{12}$CO (dashed line) and HI (dash-dotted line) spectra obtained the region highest X-ray absorption. 
}
   \label{fig:COspec}
\end{figure}

The comparison of the  absorption along the line of sight derived from X-ray data,  $^{12}$CO and HI observations can be used to constrain the distance to the SNR.
The velocity spectra of the $^{12}$CO emission \citep[using data from the CfA survey, ][]{dame01} and the HI emission 
\citep[using data from the SGPS survey, ][]{hg06}  derived from the region of highest X-ray 
absorption ($\alpha_{\mathrm{J2000}}=$17h31m43s,  $\delta_{\mathrm{J2000}}=-34\degr34'58''$) are shown in Fig.~\ref{fig:COspec} (bottom).

In order to derive a lower limit on the integration distance required to match the $N_{\rm H}$ derived from X-rays, all the material is
assumed to be at the near distance allowed by the Galactic rotation curve. 
Under this hypothesis,  the cumulative absorption column derived from the atomic and molecular  hydrogen shown in Fig.~\ref{fig:COspec} (top) 
is similar to the one observed in X-rays, $N_{\mathrm{H}} = (2.23   \pm 0.21) \times 10^{22}$ cm$^{-2}$, when integrating
 up to a radial velocity relative to the \textit{local standard of rest} (LSR) of $-$25 km/s.
 The CO-to-H$_{2}$ mass conversion factor and the HI brightness temperature to column density used are respectively of
$1.8\times10^{20}$ cm$^{-2}$ K$^{-1}$ km$^{-1}$ s \citep{dame01} and $1.82\times10^{18}$ cm$^{-2}$ K$^{-1}$ km$^{-1}$ s \citep{dickey90}.
 
 When integrating up to the same velocity, the  map of $N_{\rm H}$ derived from the atomic and molecular hydrogen shown in Fig.~\ref{fig:xraynh} 
(right) exhibits an increase of absorption towards the Galactic plane similar to that in the X-ray absorption map  in Fig.~\ref{fig:xraynh} (left).
The peak of $^{12}$CO emission at a LSR velocity of $-$18 km/s is thus in the foreground of the SNR and is likely to be the cause of the gradient of absorption seen in X-rays.
Using the circular Galactic rotation model of \cite{hou09} with a distance to the Galactic center of 8.0 kpc, the 
nearest distance corresponding to the LSR velocity of $-$18 km/s is \dist kpc thus setting a lower limit for the distance of the remnant.

\begin{figure}
   \centering
\hspace{-0.3cm}   \includegraphics[bb=56 210 525 585,clip,width=9.3cm]{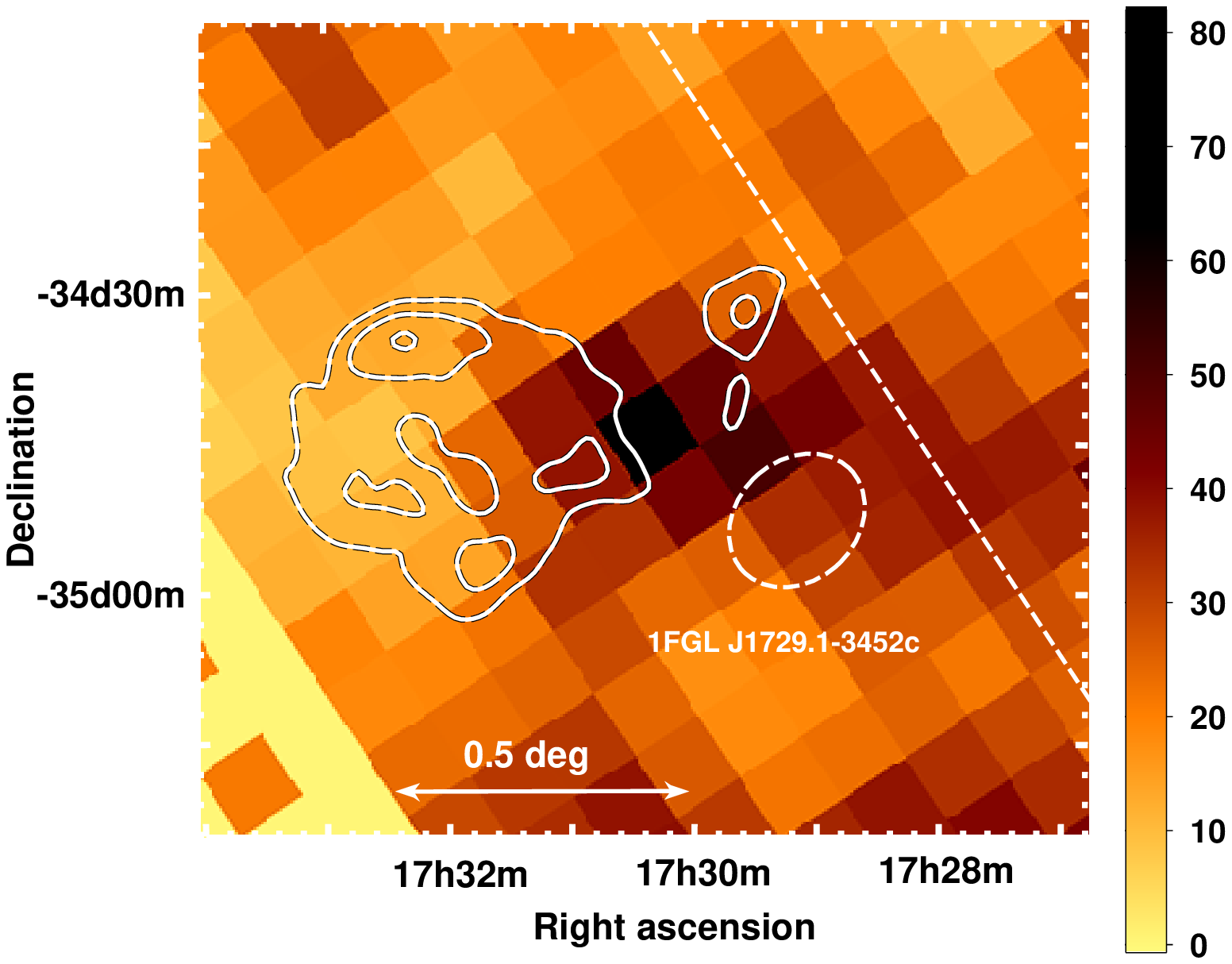}
\hspace{-0.43cm}    \includegraphics[bb=56 210 525 585,clip,width=9.3cm]{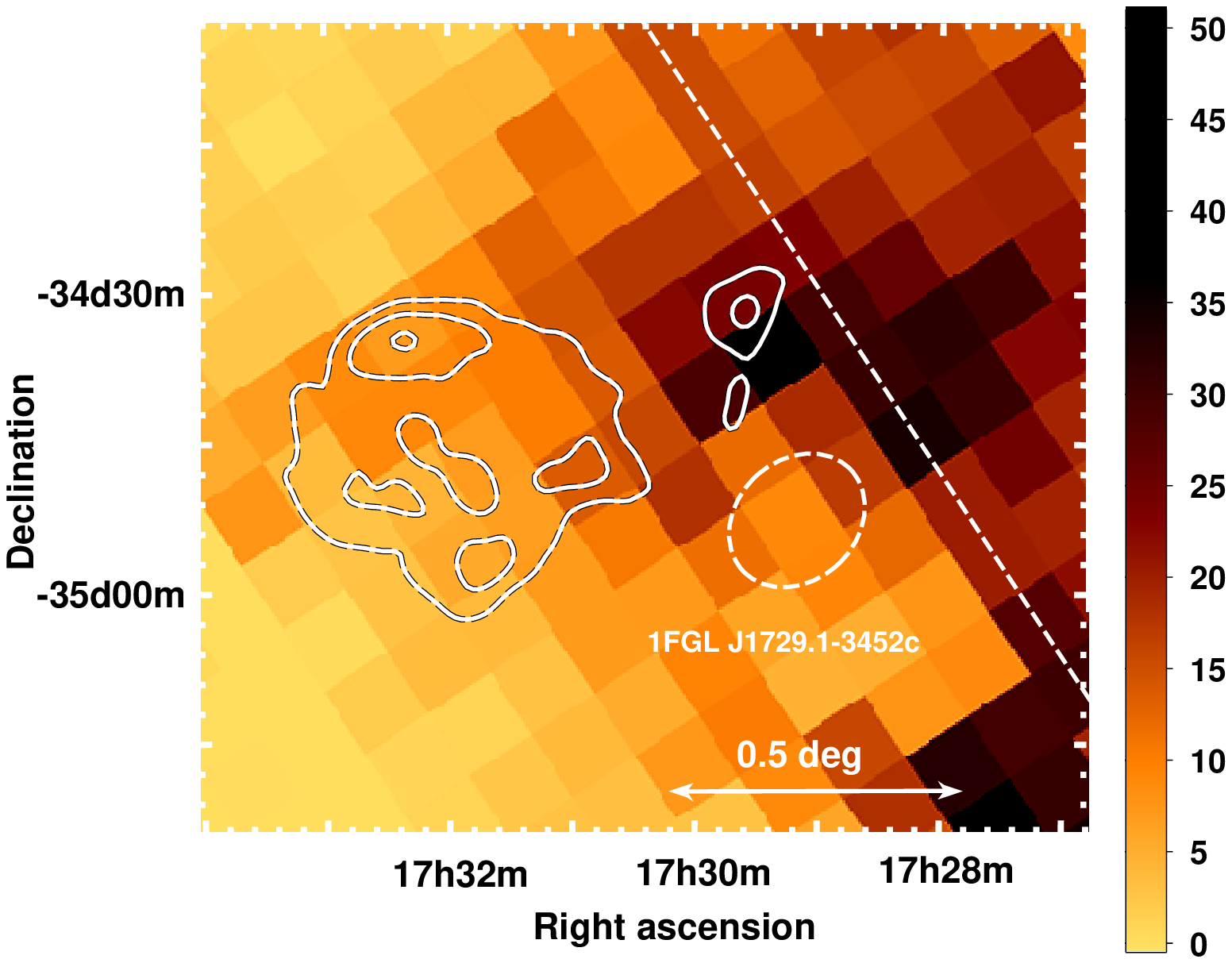}
   \caption{$^{12}$CO map of the vicinity of \src integrated from LSR velocity $-$13 km/s to $-$25 km/s (top) and from $-$75 km/s to $-$87 km/s (bottom)
    respectively corresponding to the intervals of the second and third $^{12}$CO peak shown in Fig.~\ref{fig:COspec}.
   The HESS significance contours from Fig.~\ref{fig:4contrib} together with the Galactic plane and the 
   Fermi 95\% position confidence level contours presented in the 1 year catalog are overlaid.
   The linear scale is in units of K km/s. }
   \label{fig:COslice}
\end{figure}

\subsection{GeV \g-rays}
\label{sect:fermi}

In the Fermi-LAT first year catalog \citep{abdo10cat} the source 1FGL J1729.1$-$3452c is found in the neighborhood of \src as shown in Fig.~\ref{fig:COslice}. 
The Fermi source has an analysis flag that indicates that the source position  moved beyond its 95\% error ellipse when changing the model 
of diffuse emission.
The $^{12}$CO map in Fig.~\ref{fig:COslice} shows that the Fermi source is located near a small scale gas clump that could be not well represented in the  
diffuse emission model. The position of the source presented in the catalog is to be used with caution and is therefore possibly not incompatible with \srcb.

The Fermi source has a photon spectral slope of 2.26$\pm$0.08 and shows neither indication for spectral curvature nor time variability on a time scale of months 
(the catalog does not address shorter or longer time variations). This source is the closest Fermi detection
near the newly discovered SNR and the flux derived in the Fermi catalog is used as an upper limit in the SED of the SNR in Fig.~\ref{fig:SED}.

\subsection{Multi-wavelength counterparts for \srcb}
\label{sect:robin}

At radio wavelengths, the \g-ray contours of \srcb lie near the HII region G353.381-0.114. Using HI radio recombination line data, the LSR velocity
corresponding to this source is either $-$54 km/s or $-$82 km/s \citep{caswell87}.
 In the latter case this HII region could be associated with the molecular cloud observed around
velocities of $\sim$ $-$80 km/s (see Fig.~\ref{fig:COslice}). 
At X-ray energies, no archival dedicated observations were found, and no emission is detected in the ROSAT all sky survey, probably due to the high 
absorption in the line of sight. As discussed in the previous section, a Fermi source is found to lie close to \srcb.

\section{Discussion}
\label{sect:discussion}

The newly discovered SNR \srca is in several ways comparable to \rxj and \vela. Those objects are X-ray synchrotron emitters and exhibit
 no thermal emission lines. A CCO is also found within those three SNRs indicating a core collapse SN.
Moreover at a distance of \dist kpc (see Sect. \ref{sect:CO}),
the TeV luminosity of \srca in the 1-30 TeV energy band is 1.07$\times (d/\dist \, \rm{kpc})^{2} \, 10^{34}$ erg s$^{-1}$ which is similar to the luminosity of \rxj 
(the brightest TeV shell SNR detected until now), of 0.81$\times 10^{34}$ erg s$^{-1}$ using a distance of 1 kpc \citep{fm03,cd04,mt05} 
and slightly higher than that of  \vela with 0.65$\times 10^{34}$ erg s$^{-1}$ at a distance of 0.75 kpc \citep{katsuda08}.
A difference with \rxj is that the flat \g-ray  azimuthal profile of \srca (see Fig.~\ref{fig:azprof}) suggests that the remnant is evolving in a relatively uniform ambient medium and 
that it is not in interaction with the cloud (shown in Fig.~\ref{fig:COslice}, top) used to derive a lower limit to the distance of the SNR. 
This significantly differs from the case of \rxj which exhibits much brighter \g-ray emission in the North-West where the shock is thought to interact with denser material.

The distance used for the luminosity is derived from the absorption in the foreground and provides only a lower limit of \dist kpc.
However, as it is believed that supernova explosions are more likely to  occur in the spiral arms of the Galaxy  where the density of massive stars 
(i.e. SNR progenitors) is higher \citep{russeil03,hou09}, it is likely that \srca could be located within
the  Scutum-Crux or  Norma arms, which cross the line of sight at $l=353.5^{\circ}$ at $\sim$3.0 and $\sim$4.5 kpc respectively \citep{hou09}.
The next arm in the  same line of sight is the Sagittarius arm lying at a distance of 12 kpc. This latter possibility for the location
of the SNR would lead to a much higher \g-ray luminosity,  an order of magnitude higher than \rxj. Also at such a distance, the physical size of the 
remnant would exceed 50 pc, substantially larger than other TeV shell SNRs whose physical size is $\lesssim$ 15 pc. 
As a result it is reasonable to believe that the real distance to the SNR should not be much larger than the derived lower limit of \dist kpc. 

\begin{figure}
   \centering
   \includegraphics[bb= 142 55 575 707, clip,width=6cm,angle=-90]{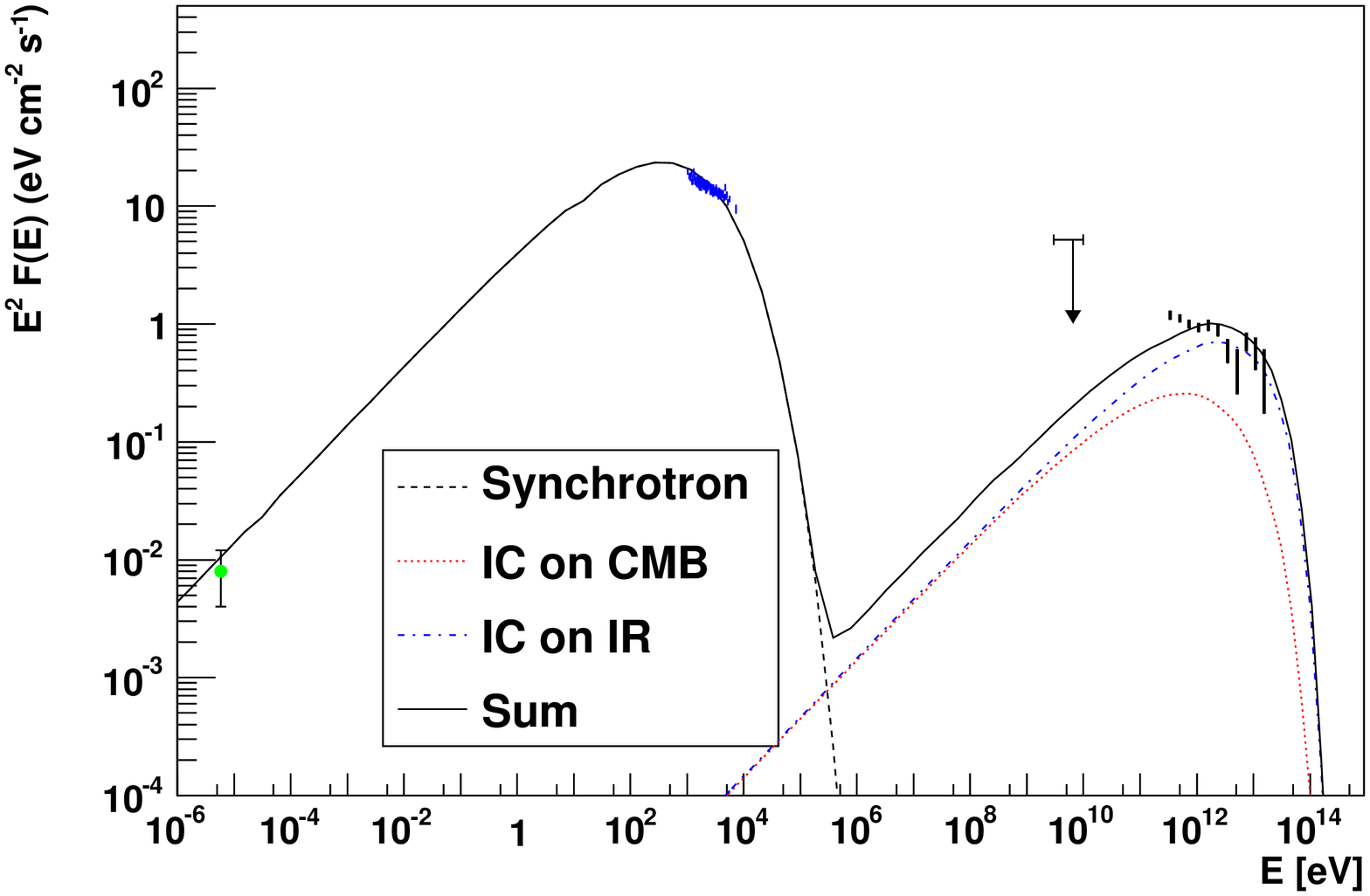} \\
   \includegraphics[bb= 142 55 575 707 , clip,width=6cm,angle=-90]{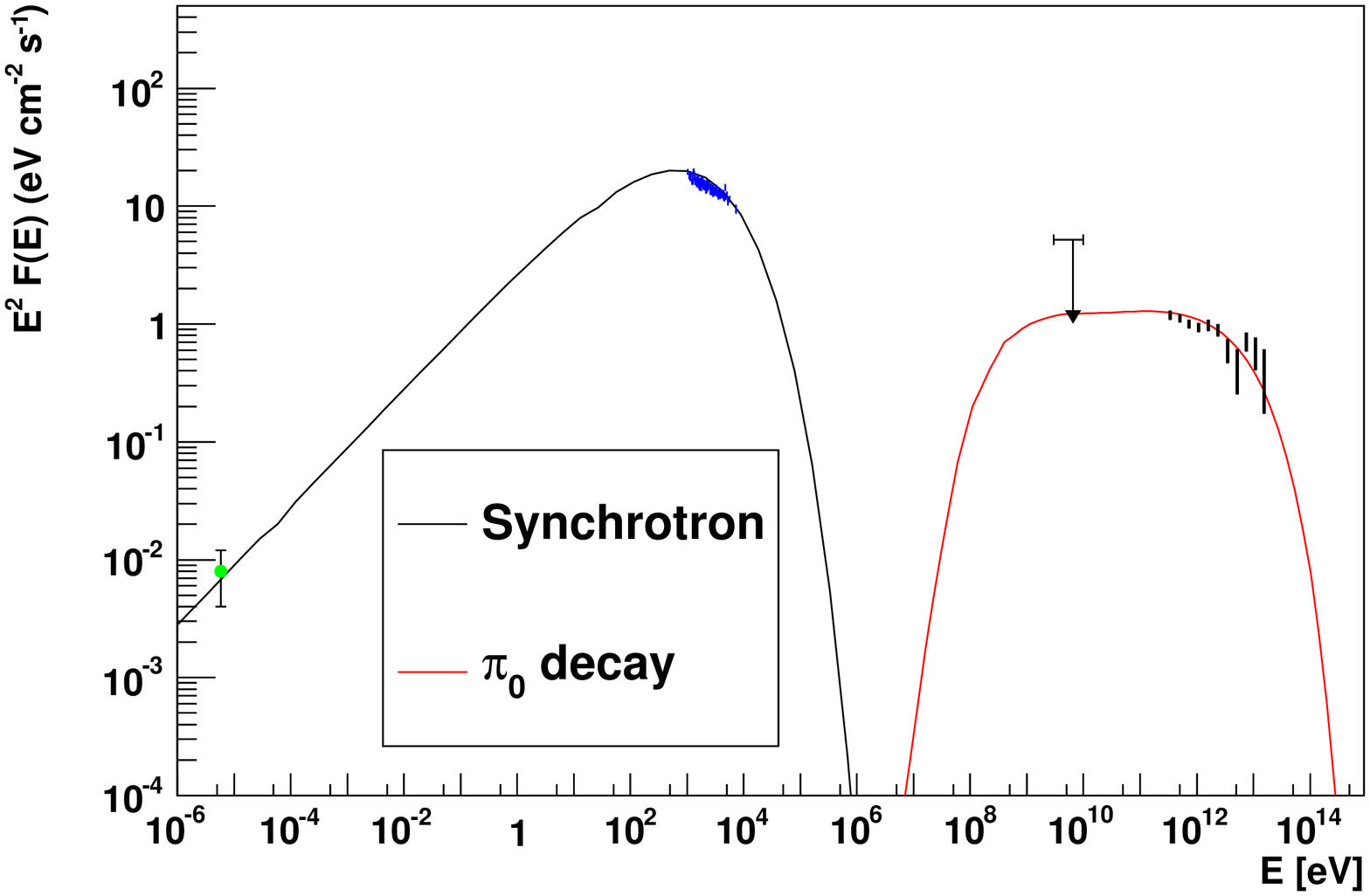} 
      \vspace{-0.2cm}

   \caption{Broadband SED for the sub-region of \srca that is observed in X-rays (see Fig.~\ref{fig:4contrib}, right, for region definition).
    A purely leptonic (top) and a hadronic (bottom) scenario  are shown and the corresponding parameters for both models are
     presented in Table \ref{tab:SED}. The infrared (IR) seed photons energy density and temperature were 
     respectively set to 1 eV cm$^{-3}$ and 40 K following the model from \citet{porter08} for a galactocentric radius
     of 4 kpc. 
     The flux from the nearby Fermi source 1FGL J1729.1$-$3452c is represented and used as an upper limit. }
   \label{fig:SED}
\end{figure}

The radio flux and the X- and \g-ray spectra derived in Sect. \ref{sect:multi} from the sub-region of \srca that is covered by the FoV of \x  were combined
in the SED presented in Fig.~\ref{fig:SED}. The X-ray data were corrected for the interstellar absorption with 
$N_{\rm H}=1.08 \times 10^{22}$ cm$^{-2}$.
 To model the SED, a simple one-zone stationary model \citep[presented in ][]{acero10} was used.  
In this model, the spectrum of electrons and protons is represented by a power-law of slope $s$ with exponential cutoffs at energies
$E_{\rm c,e}$ and $E_{\rm c,p}$ for the electrons and protons respectively.   
For the modeling of the object, it is assumed that the measured multi-wavelength emission from the sub-region of \srca is 
entirely coming from the SNR located at a distance of \dist kpc. 
As this distance is only a lower limit, the total energy of accelerated particles ($W_{\rm e}$ and $W_{\rm p}$) 
in the SNR should also be viewed as lower limits.

In the pure leptonic scenario, the slope of the electrons is constrained by the radio and the X-ray synchrotron emission between 1.9 and 2.1 and
the strength of the magnetic field required to reproduce the ratio of observed synchrotron and IC emission 
lies between 20 and 30 $\mu$G  for $15 \leq E_{\rm c,e} \leq 25$ TeV. Although the relative ratio of radio, X- and TeV fluxes can be fairly well
reproduced by this leptonic scenario, the model is inadequate to account for the X-ray and the  \g-ray spectral slope 
as illustrated in Fig.~\ref{fig:SED} (top). The corresponding parameters for the latter model are summarized in Table \ref{tab:SED}.

This limitation no longer occurs in a scenario where the TeV emission is dominated by hadronic
processes as the X- and \g-ray emission are now independent and both spectral slopes can be reproduced as shown in Fig.~\ref{fig:SED} (bottom).
Moreover, the strength of the magnetic field can be increased  as it is no longer fixed by the X/\g ratio. 
In order to reproduce the observed TeV flux, the total energy in high-energy protons ($E \, \geq$ 1 GeV) assuming a spectral slope of 2.0 
is $W_{\rm p} = 2 \times 10^{50}$ ($n$/1 cm$^{-3})^{-1}$ ($d$/\dist kpc)$^{2}$ erg. 
It should be noted that this energy content only represents a sub-region of the SNR accounting for $\sim$ 1/3 of the total TeV flux
(see Table \ref{tab:param}) implying that the total energy transferred to accelerated protons in the whole SNR is 
 a substantial fraction of the energy available in the remnant for $n \sim$ 1 cm$^{-3}$.
For this energetic reason, gas densities much below this value appear incompatible with the hadronic emission scenario.

Although it is not possible to measure the density of the ambient medium surrounding the SNR as no X-ray thermal emission is detected,
an upper limit on the density can be derived. 
In order to do so, a thermal component, whose normalization is fixed for a given density using the method presented in \citet{ab07} (Sect. 3.1), is
 added to the X-ray spectrum. 
 The shocked ambient medium is assumed to be in a non equilibrium ionization state with an ionization timescale parameter
$\tau=10^{9}$ cm$^{-3}$ s and an electron plasma
temperature $kT_{\rm e}$=1 keV.  Such values are commonly observed in other young SNRs 
for which the X-ray emission of the shocked ambient medium has been studied as in e.g. RCW86 \citep[see][]{vink06}.
For the given parameters, the derived upper limit (90\% confidence level) on the ambient medium density is $10^{-2}$ cm$^{-3}$.
In the case of a lower temperature ($kT_{\rm e}$=0.15 keV),  an upper limit of 1 cm$^{-3}$ is reached.

For a density of 1 cm$^{-3}$  the corresponding shock speed and age of the SNR
would be  $\sim$410 km/s and 14000 yrs in order to match a physical radius of 
$R_{\rm shock}$=15 pc (0.27$^{\circ}$ at \dist kpc), for  a  SN  explosion of $E_{\rm SN}=1 \times 10^{51}$ erg with a mass  of ejecta of 5 $M_\odot$ 
using equations from \citet{tm99}.
However, this shock speed is an order of magnitude lower than what has been measured in other bright synchrotron emitting SNRs like SN 1006 
\citep[$V_{\rm sh}$=5000 $\pm$ 400 km/s at a distance of 2.2 kpc ; ][]{katsuda09}, RCW 86 \citep[$V_{\rm sh}$=6000 $\pm$ 3000  km/s ; ][]{helder09}, 
CasA \citep[$V_{\rm sh}$=4900  km/s ; ][]{patnaude09} or Tycho \citep[$V_{\rm sh} = 3000 \pm 1000 $ km/s at a distance of 2.3 kpc ; ][]{katsuda10}. 
As a rough estimate, the required density  to reproduce a canonical shock speed of 3000 km/s  using the aforementioned SN parameters
is of the order of 0.01 cm$^{-3}$  (compatible with the upper limit derived from the lack of thermal X-ray emission in the previous paragraph) for a corresponding age of $\sim$ 2500 yrs. 

To summarize, the presented static one-zone model suffers from limitations in both scenarios. In the leptonic case,
the model allows to estimate  the average B-field ($\sim 25 \, \mu$G) and the total energy in accelerated electrons present in the shell of the SNR but
 fails to reproduce the observed X-ray and \g-ray spectral slope. In the hadronic model, the high medium density required to 
reproduce the observed TeV flux is hardly compatible with the hydrodynamics of the SNR.
More detailed models using  non-linear diffuse shock acceleration theory have been developed 
\citep[e.g.][]{zirakashvili10,ellison10} and would  provide more accurate predictions than the simple model presented here.
It should be noted that the considered model does not take into account evolution related to radiative cooling
which  could yield a steeper gamma-ray spectrum, in better agreement with the data.  
Also, the presented scenarios do not cover possible non-homogeneous surroundings such as wind bubble blown by the progenitor.
Such detailed spectral and evolutionary modeling depends on many poorly known parameters and is therefore beyond the scope of the
present discussion.

Concerning the source \srcb, detected in the vicinity of the SNR, the presented multi-wavelength data
do not provide a clear understanding of the nature of the object. 
The closest structures located near the \g-ray emission are 
the  HII region G353.381-0.114 (seen in the radio in Fig.~\ref{fig:4contrib}, left) and
a molecular gas clump observed in $^{12}$CO (see Fig.~\ref{fig:COslice}, bottom) when integrating around a LSR velocity
 of $-$80 km/s  (corresponding to near and far kinematic distances of $\sim$6 and $\sim$10 kpc respectively). 
If the \g-ray source \srcb is associated with those gas structures, it would therefore not be associated with the SNR HESS J1731-347 thought to lie at a closer distance.

\begin{table}
\centering
\caption{List of the parameters used for the spectral energy distribution modeling presented in Fig. \ref{fig:SED}. 
The spectral slope are  fixed at 2.0 for the electron and the proton distribution. The density of the ambient medium was set to 1 cm$^{-3}$ in the case of the 
hadronic model.}
\label{tab:SED}

\renewcommand{\footnoterule}{}  
\begin{tabular}{l c c c c c }     
\hline

Model & $E_{\rm c,e}$ & $E_{\rm c,p}$ & $W_{\rm e}$ & $W_{\rm p}$ & $B$ \\
           &          TeV                &  TeV            &   $10^{47}$ erg  & $10^{50}$ erg & $\mu$G \\
\hline

leptonic   &        18     &       $-$         &      1.1      &       $-$       &      25    \\
hadronic   &       16      &        100        &       0.25     &       2.0       &    50     \\

\hline
\end{tabular}
\end{table}

\section{Conclusion}

The newly discovered SNR \srca exhibits a significant shell morphology spatially resolved by \h,  similar to the one observed in radio.
Together with \rxj, \vela and SN 1006, \srca is now the fourth\footnote{With the current statistics, the shell morphology of RCW86 is not statistically significant \citep{aharonian09}.} TeV \g-ray source to join this small but growing class. 
A lower limit to the distance of the SNR of \dist kpc was obtained by comparing the absorption derived from the X-rays and from HI and $^{12}$CO observations.

The multi-wavelength emission from 
the SNR, detected in radio, X-rays and  \g-rays, was combined in an SED to investigate the origin of the \g-ray emission assuming that the broadband emission
stems from the same region (one-zone model). While the measured fluxes can
be accounted for in a purely leptonic model with a magnetic field of the order of 25 $\mu$G, this simple model fails to reproduce the spectral shape of
 the X- and  \g-ray emission.
A second model that assumes that the TeV emission is produced by hadronic processes is able to reproduce the spectral slopes in X- and \g-rays at the cost
of requiring that a large fraction of the kinetic energy of the explosion must be transferred to the accelerated protons and a high ambient medium density 
of $n\sim$1  cm$^{-3}$ for $d \geq$ \dist kpc.
Moreover  for such a density, the corresponding shock speed of the SNR would be an order of magnitude lower than in other SNRs exhibiting bright 
synchrotron emission.

\begin{acknowledgements}
The support of the Namibian authorities and of the University of Namibia
in facilitating the construction and operation of HESS is gratefully
acknowledged, as is the support by the German Ministry for Education and
Research (BMBF), the Max Planck Society, the French Ministry for Research,
the CNRS-IN2P3 and the Astroparticle Interdisciplinary Programme of the
CNRS, the U.K. Science and Technology Facilities Council (STFC),
the IPNP of the Charles University, the Polish Ministry of Science and 
Higher Education, the South African Department of
Science and Technology and National Research Foundation, and by the
University of Namibia. We appreciate the excellent work of the technical
support staff in Berlin, Durham, Hamburg, Heidelberg, Palaiseau, Paris,
Saclay, and in Namibia in the construction and operation of the
equipment. Article based on observations obtained with XMM-Newton, 
an ESA science mission with instruments and contributions directly 
funded by ESA Member States and NASA. 
\end{acknowledgements}


\vspace{-0.4cm}



\begin{thebibliography}{44}
\expandafter\ifx\csname natexlab\endcsname\relax\def\natexlab#1{#1}\fi

\bibitem[{{Abdo} {et~al.}(2010){Abdo}, {Ackermann}, {Ajello}, {Allafort},
  {Antolini}, {Atwood}, {Axelsson}, {Baldini}, {Ballet}, {Barbiellini}, \&
  et~al.}]{abdo10cat}
{Abdo}, A.~A., {Ackermann}, M., {Ajello}, M., {et~al.} 2010, \apjs, 188, 405

\bibitem[{{Acero} {et~al.}(2010){Acero}, {Aharonian}, {Akhperjanian}, {Anton},
  {Barres de Almeida}, {Bazer-Bachi}, {Becherini}, {Behera}, {Beilicke},
  {Bernl{\"o}hr}, {Bochow}, {Boisson}, {Bolmont}, {Borrel}, {Brucker}, {Brun},
  {Brun}, {B{\"u}hler}, {Bulik}, {B{\"u}sching}, {Boutelier}, {Chadwick},
  {Charbonnier}, {Chaves}, {Cheesebrough}, {Conrad}, {Chounet}, {Clapson},
  {Coignet}, {Dalton}, {Daniel}, {Davids}, {Degrange}, {Deil}, {Dickinson},
  {Djannati-Ata{\"\i}}, {Domainko}, {O'C.~Drury}, {Dubois}, {Dubus}, {Dyks},
  {Dyrda}, {Egberts}, {Eger}, {Espigat}, {Fallon}, {Farnier}, {Fegan},
  {Feinstein}, {Fiasson}, {F{\"o}rster}, {Fontaine}, {F{\"u}{\ss}ling},
  {Gabici}, {Gallant}, {G{\'e}rard}, {Gerbig}, {Giebels}, {Glicenstein},
  {Gl{\"u}ck}, {Goret}, {G{\"o}ring}, {Hauser}, {Hauser}, {Heinz},
  {Heinzelmann}, {Henri}, {Hermann}, {Hinton}, {Hoffmann}, {Hofmann},
  {Hofverberg}, {Holleran}, {Hoppe}, {Horns}, {Jacholkowska}, {de Jager},
  {Jahn}, {Jung}, {Katarzy{\'n}ski}, {Katz}, {Kaufmann}, {Kerschhaggl},
  {Khangulyan}, {Kh{\'e}lifi}, {Keogh}, {Klochkov}, {Klu{\'z}niak}, {Kneiske},
  {Komin}, {Kosack}, {Kossakowski}, {Lamanna}, {Lemoine-Goumard}, {Lenain},
  {Lohse}, {Marandon}, {Marcowith}, {Masbou}, {Maurin}, {McComb}, {Medina},
  {M{\'e}hault}, {Moderski}, {Moulin}, {Naumann-Godo}, {de Naurois}, {Nedbal},
  {Nekrassov}, {Nicholas}, {Niemiec}, {Nolan}, {Ohm}, {Olive}, {de O{\~n}a
  Wilhelmi}, {Orford}, {Ostrowski}, {Panter}, {Paz Arribas}, {Pedaletti},
  {Pelletier}, {Petrucci}, {Pita}, {P{\"u}hlhofer}, {Punch}, {Quirrenbach},
  {Raubenheimer}, {Raue}, {Rayner}, {Reimer}, {Renaud}, {de Los Reyes},
  {Rieger}, {Ripken}, {Rob}, {Rosier-Lees}, {Rowell}, {Rudak}, {Rulten},
  {Ruppel}, {Ryde}, {Sahakian}, {Santangelo}, {Schlickeiser}, {Sch{\"o}ck},
  {Sch{\"o}nwald}, {Schwanke}, {Schwarzburg}, {Schwemmer}, {Shalchi}, {Sushch},
  {Sikora}, {Skilton}, {Sol}, {Stawarz}, {Steenkamp}, {Stegmann}, {Stinzing},
  {Superina}, {Szostek}, {Tam}, {Tavernet}, {Terrier}, {Tibolla}, {Tluczykont},
  {van Eldik}, {Vasileiadis}, {Venter}, {Venter}, {Vialle}, {Vincent}, {Vink},
  {Vivier}, {V{\"o}lk}, {Volpe}, {Vorobiov}, {Wagner}, {Ward}, {Zdziarski}, \&
  {Zech}}]{acero10}
{Acero}, F., {Aharonian}, F., {Akhperjanian}, A.~G., {et~al.} 2010, \aap, 516,
  A62

\bibitem[{{Acero} {et~al.}(2007){Acero}, {Ballet}, \& {Decourchelle}}]{ab07}
{Acero}, F., {Ballet}, J., \& {Decourchelle}, A. 2007, \aap, 475, 883

\bibitem[{{Acero} {et~al.}(2009{\natexlab{a}}){Acero}, {Ballet},
  {Decourchelle}, {Lemoine-Goumard}, {Ortega}, {Giacani}, {Dubner}, \&
  {Cassam-Chena{\"\i}}}]{acero09}
{Acero}, F., {Ballet}, J., {Decourchelle}, A., {et~al.} 2009{\natexlab{a}},
  \aap, 505, 157

\bibitem[{{Acero} {et~al.}(2009{\natexlab{b}}){Acero}, {P{\"u}hlhofer},
  {Klochkov}, {Komin}, {Gallant}, {Horns}, {Santangelo}, \& {for the
  H.~E.~S.~S.~Collaboration}}]{ap09}
{Acero}, F., {P{\"u}hlhofer}, G., {Klochkov}, D., {et~al.} 2009{\natexlab{b}},
  ArXiv e-prints 0907.0642

\bibitem[{{Aharonian} {et~al.}(2008){Aharonian}, {Akhperjanian}, {Barres de
  Almeida}, {Bazer-Bachi}, {Behera}, {Beilicke}, {Benbow}, {Bernl{\"o}hr},
  {Boisson}, {Bolz}, {Borrel}, {Braun}, {Brion}, {Brown}, {B{\"u}hler},
  {Bulik}, {B{\"u}sching}, {Boutelier}, {Carrigan}, {Chadwick}, {Chounet},
  {Clapson}, {Coignet}, {Cornils}, {Costamante}, {Dalton}, {Degrange},
  {Dickinson}, {Djannati-Ata{\"\i}}, {Domainko}, {Drury}, {Dubois}, {Dubus},
  {Dyks}, {Egberts}, {Emmanoulopoulos}, {Espigat}, {Farnier}, {Feinstein},
  {Fiasson}, {F{\"o}rster}, {Fontaine}, {Funk}, {F{\"u}{\ss}ling}, {Gallant},
  {Giebels}, {Glicenstein}, {Gl{\"u}ck}, {Goret}, {Hadjichristidis}, {Hauser},
  {Hauser}, {Heinzelmann}, {Henri}, {Hermann}, {Hinton}, {Hoffmann}, {Hofmann},
  {Holleran}, {Hoppe}, {Horns}, {Jacholkowska}, {de Jager}, {Jung},
  {Katarzy{\'n}ski}, {Kendziorra}, {Kerschhaggl}, {Kh{\'e}lifi}, {Keogh},
  {Komin}, {Kosack}, {Lamanna}, {Latham}, {Lemi{\`e}re}, {Lemoine-Goumard},
  {Lenain}, {Lohse}, {Martin}, {Martineau-Huynh}, {Marcowith}, {Masterson},
  {Maurin}, {Maurin}, {McComb}, {Moderski}, {Moulin}, {de Naurois}, {Nedbal},
  {Nolan}, {Ohm}, {Olive}, {de O{\~n}a Wilhelmi}, {Orford}, {Osborne},
  {Ostrowski}, {Panter}, {Pedaletti}, {Pelletier}, {Petrucci}, {Pita},
  {P{\"u}hlhofer}, {Punch}, {Ranchon}, {Raubenheimer}, {Raue}, {Rayner},
  {Renaud}, {Ripken}, {Rob}, {Rolland}, {Rosier-Lees}, {Rowell}, {Rudak},
  {Ruppel}, {Sahakian}, {Santangelo}, {Schlickeiser}, {Sch{\"o}ck},
  {Schr{\"o}der}, {Schwanke}, {Schwarzburg}, {Schwemmer}, {Shalchi}, {Sol},
  {Spangler}, {Stawarz}, {Steenkamp}, {Stegmann}, {Superina}, {Tam},
  {Tavernet}, {Terrier}, {van Eldik}, {Vasileiadis}, {Venter}, {Vialle},
  {Vincent}, {Vivier}, {V{\"o}lk}, {Volpe}, {Wagner}, {Ward}, {Zdziarski}, \&
  {Zech}}]{ah08}
{Aharonian}, F., {Akhperjanian}, A.~G., {Barres de Almeida}, U., {et~al.} 2008,
  \aap, 477, 353

\bibitem[{{Aharonian} {et~al.}(2006){Aharonian}, {Akhperjanian}, {Bazer-Bachi},
  {Beilicke}, {Benbow}, {Berge}, {Bernl{\"o}hr}, {Boisson}, {Bolz}, {Borrel},
  {Braun}, {Breitling}, {Brown}, {Chadwick}, {Chounet}, {Cornils},
  {Costamante}, {Degrange}, {Dickinson}, {Djannati-Ata{\"\i}}, {Drury},
  {Dubus}, {Emmanoulopoulos}, {Espigat}, {Feinstein}, {Fontaine}, {Fuchs},
  {Funk}, {Gallant}, {Giebels}, {Gillessen}, {Glicenstein}, {Goret},
  {Hadjichristidis}, {Hauser}, {Heinzelmann}, {Henri}, {Hermann}, {Hinton},
  {Hofmann}, {Holleran}, {Horns}, {Jacholkowska}, {de Jager}, {Kh{\'e}lifi},
  {Komin}, {Konopelko}, {Latham}, {Le Gallou}, {Lemi{\`e}re},
  {Lemoine-Goumard}, {Leroy}, {Lohse}, {Martin}, {Martineau-Huynh},
  {Marcowith}, {Masterson}, {McComb}, {de Naurois}, {Nolan}, {Noutsos},
  {Orford}, {Osborne}, {Ouchrif}, {Panter}, {Pelletier}, {Pita},
  {P{\"u}hlhofer}, {Punch}, {Raubenheimer}, {Raue}, {Raux}, {Rayner}, {Reimer},
  {Reimer}, {Ripken}, {Rob}, {Rolland}, {Rowell}, {Sahakian}, {Saug{\'e}},
  {Schlenker}, {Schlickeiser}, {Schuster}, {Schwanke}, {Siewert}, {Sol},
  {Spangler}, {Steenkamp}, {Stegmann}, {Tavernet}, {Terrier}, {Th{\'e}oret},
  {Tluczykont}, {Vasileiadis}, {Venter}, {Vincent}, {V{\"o}lk}, \&
  {Wagner}}]{aharonian06}
{Aharonian}, F., {Akhperjanian}, A.~G., {Bazer-Bachi}, A.~R., {et~al.} 2006,
  \apj, 636, 777

\bibitem[{{Aharonian} {et~al.}(2009){Aharonian}, {Akhperjanian}, {de Almeida},
  {Bazer-Bachi}, {Behera}, {Beilicke}, {Benbow}, {Bernl{\"o}hr}, {Boisson},
  {Bochow}, {Borrel}, {Braun}, {Brion}, {Brucker}, {B{\"u}hler}, {Bulik},
  {B{\"u}sching}, {Boutelier}, {Carrigan}, {Chadwick}, {Charbonnier}, {Chaves},
  {Chounet}, {Clapson}, {Coignet}, {Costamante}, {Dalton}, {Degrange},
  {Dickinson}, {Djannati-Ata{\"\i}}, {Domainko}, {Drury}, {Dubois}, {Dubus},
  {Dyks}, {Egberts}, {Emmanoulopoulos}, {Espigat}, {Farnier}, {Feinstein},
  {Fiasson}, {F{\"o}rster}, {Fontaine}, {F{\"u}{\ss}ling}, {Gabici}, {Gallant},
  {G{\'e}rard}, {Giebels}, {Glicenstein}, {Gl{\"u}ck}, {Goret},
  {Hadjichristidis}, {Hauser}, {Hauser}, {Heinzelmann}, {Henri}, {Hermann},
  {Hinton}, {Hoffmann}, {Hofmann}, {Holleran}, {Hoppe}, {Horns},
  {Jacholkowska}, {de Jager}, {Jung}, {Katarzy{\'n}ski}, {Kaufmann},
  {Kendziorra}, {Kerschhaggl}, {Khangulyan}, {Kh{\'e}lifi}, {Keogh}, {Komin},
  {Kosack}, {Lamanna}, {Latham}, {Lemoine-Goumard}, {Lenain}, {Lohse},
  {Marandon}, {Martin}, {Martineau-Huynh}, {Marcowith}, {Masterson}, {Maurin},
  {McComb}, {Medina}, {Moderski}, {Moulin}, {Naumann-Godo}, {de Naurois},
  {Nedbal}, {Nekrassov}, {Niemiec}, {Nolan}, {Ohm}, {Olive}, {de O{\~n}a
  Wilhelmi}, {Orford}, {Osborne}, {Ostrowski}, {Panter}, {Pedaletti},
  {Pelletier}, {Petrucci}, {Pita}, {P{\"u}hlhofer}, {Punch}, {Quirrenbach},
  {Raubenheimer}, {Raue}, {Rayner}, {Renaud}, {Rieger}, {Ripken}, {Rob},
  {Rosier-Lees}, {Rowell}, {Rudak}, {Ruppel}, {Sahakian}, {Santangelo},
  {Schlickeiser}, {Sch{\"o}ck}, {Schr{\"o}der}, {Schwanke}, {Schwarzburg},
  {Schwemmer}, {Shalchi}, {Skilton}, {Sol}, {Spangler}, {Stawarz}, {Steenkamp},
  {Stegmann}, {Superina}, {Tam}, {Tavernet}, {Terrier}, {Tibolla}, {van Eldik},
  {Vasileiadis}, {Venter}, {Vialle}, {Vincent}, {Vink}, {Vivier}, {V{\"o}lk},
  {Volpe}, {Wagner}, {Ward}, {Zdziarski}, \& {Zech}}]{aharonian09}
{Aharonian}, F., {Akhperjanian}, A.~G., {de Almeida}, U.~B., {et~al.} 2009,
  \apj, 692, 1500

\bibitem[{{Berge} {et~al.}(2007){Berge}, {Funk}, \& {Hinton}}]{berge07}
{Berge}, D., {Funk}, S., \& {Hinton}, J. 2007, \aap, 466, 1219

\bibitem[{{Bernl{\"o}hr} {et~al.}(2003){Bernl{\"o}hr}, {Carrol}, {Cornils},
  {Elfahem}, {Espigat}, {Gillessen}, {Heinzelmann}, {Hermann}, {Hofmann},
  {Horns}, {Jung}, {Kankanyan}, {Katona}, {Khelifi}, {Krawczynski}, {Panter},
  {Punch}, {Rayner}, {Rowell}, {Tluczykont}, \& {van Staa}}]{bernlohr03}
{Bernl{\"o}hr}, K., {Carrol}, O., {Cornils}, R., {et~al.} 2003, Astroparticle
  Physics, 20, 111

\bibitem[{{Blondin} {et~al.}(2001){Blondin}, {Chevalier}, \&
  {Frierson}}]{blondin01}
{Blondin}, J.~M., {Chevalier}, R.~A., \& {Frierson}, D.~M. 2001, \apj, 563, 806

\bibitem[{{Carter} \& {Read}(2007)}]{cr07}
{Carter}, J.~A. \& {Read}, A.~M. 2007, \aap, 464, 1155

\bibitem[{{Cassam-Chena{\"\i}} {et~al.}(2004){Cassam-Chena{\"\i}},
  {Decourchelle}, {Ballet}, {Sauvageot}, {Dubner}, \& {Giacani}}]{cd04}
{Cassam-Chena{\"\i}}, G., {Decourchelle}, A., {Ballet}, J., {et~al.} 2004,
  \aap, 427, 199

\bibitem[{{Caswell} \& {Haynes}(1987)}]{caswell87}
{Caswell}, J.~L. \& {Haynes}, R.~F. 1987, \aap, 171, 261

\bibitem[{{Dame} {et~al.}(2001){Dame}, {Hartmann}, \& {Thaddeus}}]{dame01}
{Dame}, T.~M., {Hartmann}, D., \& {Thaddeus}, P. 2001, \apj, 547, 792

\bibitem[{{de Naurois} \& {Rolland}(2009)}]{dr09}
{de Naurois}, M. \& {Rolland}, L. 2009, Astroparticle Physics, 32, 231

\bibitem[{{Dickey} \& {Lockman}(1990)}]{dickey90}
{Dickey}, J.~M. \& {Lockman}, F.~J. 1990, \araa, 28, 215

\bibitem[{{Ellison} {et~al.}(2010){Ellison}, {Patnaude}, {Slane}, \&
  {Raymond}}]{ellison10}
{Ellison}, D.~C., {Patnaude}, D.~J., {Slane}, P., \& {Raymond}, J. 2010, \apj,
  712, 287

\bibitem[{{Fukui} {et~al.}(2003){Fukui}, {Moriguchi}, {Tamura}, {Yamamoto},
  {Tawara}, {Mizuno}, {Onishi}, {Mizuno}, {Uchiyama}, {Hiraga}, {Takahashi},
  {Yamashita}, \& {Ikeuchi}}]{fm03}
{Fukui}, Y., {Moriguchi}, Y., {Tamura}, K., {et~al.} 2003, \pasj, 55, L61

\bibitem[{{Funk} {et~al.}(2007{\natexlab{a}}){Funk}, {Hinton}, {Moriguchi},
  {Aharonian}, {Fukui}, {Hofmann}, {Horns}, {P{\"u}hlhofer}, {Reimer},
  {Rowell}, {Terrier}, {Vink}, \& {Wagner}}]{funk07}
{Funk}, S., {Hinton}, J.~A., {Moriguchi}, Y., {et~al.} 2007{\natexlab{a}},
  \aap, 470, 249

\bibitem[{{Funk} {et~al.}(2007{\natexlab{b}}){Funk}, {Hinton}, {P{\"u}hlhofer},
  {Aharonian}, {Hofmann}, {Reimer}, \& {Wagner}}]{funk07b}
{Funk}, S., {Hinton}, J.~A., {P{\"u}hlhofer}, G., {et~al.} 2007{\natexlab{b}},
  \apj, 662, 517

\bibitem[{{Gallant} {et~al.}(2008){Gallant}, {Carrigan}, {Djannati-Ata{\"\i}},
  {Funk}, {Hinton}, {Hoppe}, {de Jager}, {Kh{\'e}lifi}, {Komin}, {Kosack},
  {Lemi{\`e}re}, \& {Masterson}}]{gallant08}
{Gallant}, Y.~A., {Carrigan}, S., {Djannati-Ata{\"\i}}, A., {et~al.} 2008, in
  American Institute of Physics Conference Series, Vol. 983, 40 Years of
  Pulsars: Millisecond Pulsars, Magnetars and More, ed. {C.~Bassa, Z.~Wang,
  A.~Cumming, \& V.~M.~Kaspi}, 195--199

\bibitem[{{Gotthelf} \& {Halpern}(2009)}]{gotthelf09}
{Gotthelf}, E.~V. \& {Halpern}, J.~P. 2009, \apjl, 700, L158

\bibitem[{{Halpern} \& {Gotthelf}(2010)}]{halpern10}
{Halpern}, J.~P. \& {Gotthelf}, E.~V. 2010, \apj, 710, 941

\bibitem[{{Haverkorn} {et~al.}(2006){Haverkorn}, {Gaensler},
  {McClure-Griffiths}, {Dickey}, \& {Green}}]{hg06}
{Haverkorn}, M., {Gaensler}, B.~M., {McClure-Griffiths}, N.~M., {Dickey},
  J.~M., \& {Green}, A.~J. 2006, \apjs, 167, 230

\bibitem[{{Helder} {et~al.}(2009){Helder}, {Vink}, {Bassa}, {Bamba}, {Bleeker},
  {Funk}, {Ghavamian}, {van der Heyden}, {Verbunt}, \& {Yamazaki}}]{helder09}
{Helder}, E.~A., {Vink}, J., {Bassa}, C.~G., {et~al.} 2009, Science, 325, 719

\bibitem[{{Hou} {et~al.}(2009){Hou}, {Han}, \& {Shi}}]{hou09}
{Hou}, L.~G., {Han}, J.~L., \& {Shi}, W.~B. 2009, \aap, 499, 473

\bibitem[{{Katsuda} {et~al.}(2010){Katsuda}, {Petre}, {Hughes}, {Hwang},
  {Yamaguchi}, {Hayato}, {Mori}, \& {Tsunemi}}]{katsuda10}
{Katsuda}, S., {Petre}, R., {Hughes}, J.~P., {et~al.} 2010, \apj, 709, 1387

\bibitem[{{Katsuda} {et~al.}(2009){Katsuda}, {Petre}, {Long}, {Reynolds},
  {Winkler}, {Mori}, \& {Tsunemi}}]{katsuda09}
{Katsuda}, S., {Petre}, R., {Long}, K.~S., {et~al.} 2009, \apjl, 692, L105

\bibitem[{{Katsuda} {et~al.}(2008){Katsuda}, {Tsunemi}, \& {Mori}}]{katsuda08}
{Katsuda}, S., {Tsunemi}, H., \& {Mori}, K. 2008, \apjl, 678, L35

\bibitem[{{Lazendic} {et~al.}(2004){Lazendic}, {Slane}, {Gaensler}, {Reynolds},
  {Plucinsky}, \& {Hughes}}]{lazendic04}
{Lazendic}, J.~S., {Slane}, P.~O., {Gaensler}, B.~M., {et~al.} 2004, \apj, 602,
  271

\bibitem[{{Moriguchi} {et~al.}(2005){Moriguchi}, {Tamura}, {Tawara}, {Sasago},
  {Yamaoka}, {Onishi}, \& {Fukui}}]{mt05}
{Moriguchi}, Y., {Tamura}, K., {Tawara}, Y., {et~al.} 2005, \apj, 631, 947

\bibitem[{{Ohm} {et~al.}(2009){Ohm}, {van Eldik}, \& {Egberts}}]{ohm09}
{Ohm}, S., {van Eldik}, C., \& {Egberts}, K. 2009, Astroparticle Physics, 31,
  383

\bibitem[{{Patnaude} \& {Fesen}(2009)}]{patnaude09}
{Patnaude}, D.~J. \& {Fesen}, R.~A. 2009, \apj, 697, 535

\bibitem[{{Pavlov} {et~al.}(2004){Pavlov}, {Sanwal}, \& {Teter}}]{ps04}
{Pavlov}, G.~G., {Sanwal}, D., \& {Teter}, M.~A. 2004, in IAU Symposium, Vol.
  218, Young Neutron Stars and Their Environments, ed. {F.~Camilo \&
  B.~M.~Gaensler}, 239

\bibitem[{{Piron} {et~al.}(2001){Piron}, {Djannati-Atai}, {Punch}, {Tavernet},
  {Barrau}, {Bazer-Bachi}, {Chounet}, {Debiais}, {Degrange}, {Dezalay},
  {Espigat}, {Fabre}, {Fleury}, {Fontaine}, {Goret}, {Gouiffes}, {Khelifi},
  {Malet}, {Masterson}, {Mohanty}, {Nuss}, {Renault}, {Rivoal}, {Rob}, \&
  {Vorobiov}}]{piron01}
{Piron}, F., {Djannati-Atai}, A., {Punch}, M., {et~al.} 2001, \aap, 374, 895

\bibitem[{{Porter} {et~al.}(2008){Porter}, {Moskalenko}, {Strong}, {Orlando},
  \& {Bouchet}}]{porter08}
{Porter}, T.~A., {Moskalenko}, I.~V., {Strong}, A.~W., {Orlando}, E., \&
  {Bouchet}, L. 2008, \apj, 682, 400

\bibitem[{{Pratt} \& {Arnaud}(2002)}]{pa02}
{Pratt}, G.~W. \& {Arnaud}, M. 2002, \aap, 394, 375

\bibitem[{{Russeil}(2003)}]{russeil03}
{Russeil}, D. 2003, \aap, 397, 133

\bibitem[{{Tian} {et~al.}(2008){Tian}, {Leahy}, {Haverkorn}, \& {Jiang}}]{tl08}
{Tian}, W.~W., {Leahy}, D.~A., {Haverkorn}, M., \& {Jiang}, B. 2008, \apjl,
  679, L85

\bibitem[{{Tian} {et~al.}(2010){Tian}, {Li}, {Leahy}, {Yang}, {Yang},
  {Yamazaki}, \& {Lu}}]{tl10}
{Tian}, W.~W., {Li}, Z., {Leahy}, D.~A., {et~al.} 2010, \apj, 712, 790

\bibitem[{{Truelove} \& {McKee}(1999)}]{tm99}
{Truelove}, J.~K. \& {McKee}, C.~F. 1999, \apjs, 120, 299

\bibitem[{{Vink} {et~al.}(2006){Vink}, {Bleeker}, {van der Heyden}, {Bykov},
  {Bamba}, \& {Yamazaki}}]{vink06}
{Vink}, J., {Bleeker}, J., {van der Heyden}, K., {et~al.} 2006, \apjl, 648, L33

\bibitem[{{Zirakashvili} \& {Aharonian}(2010)}]{zirakashvili10}
{Zirakashvili}, V.~N. \& {Aharonian}, F.~A. 2010, \apj, 708, 965

\end{thebibliography}
\end{document}